\documentclass[amsmath,amssymb,aps,prl,lengthcheck]{revtex4-1}
\usepackage{graphicx}
\usepackage{lipsum}
\begin{document}

%\preprint{}

\title{Enhancing Light-Atom Interactions via Atomic Bunching}
\author{Bonnie L. Schmittberger and Daniel J. Gauthier}
\affiliation{Duke University, Department of Physics and Fitzpatrick Institute for Photonics, Durham, North Carolina 27708, USA}

\begin{abstract}
There is a broad interest in enhancing the strength of light-atom interactions to the point where injecting a single photon induces a nonlinear material response. Here, we show theoretically that sub-Doppler-cooled, two-level atoms that are spatially organized by weak optical fields give rise to a nonlinear material response that is greatly enhanced beyond that attainable in a homogeneous gas. Specifically, in the regime where the intensity of the applied optical fields is much less than the off-resonant saturation intensity, we show that the third-order nonlinear susceptibility scales inversely with atomic temperature and, due to this scaling, can be two orders of magnitude larger than that of a homogeneous gas for typical experimental parameters. As a result, we predict that spatially bunched two-level atoms can exhibit single-photon nonlinearities. Our model is valid for all atomic temperature regimes and simultaneously accounts for the back-action of the atoms on the optical fields. Our results agree with previous theoretical and experimental results for light-atom interactions that have considered only a limited range of temperatures. For lattice beams tuned to the low-frequency side of the atomic transition, we find that the nonlinearity transitions from a self-focusing type to a self-defocusing type at a critical intensity. We also show that higher than third-order nonlinear optical susceptibilities are significant in the regime where the dipole potential energy is on the order of the atomic thermal energy. We therefore find that it is crucial to retain high-order nonlinearities to accurately predict interactions of laser fields with spatially organized ultracold atoms. The model presented here is a foundation for modeling low-light-level nonlinear optical processes for ultracold atoms in optical lattices.
\end{abstract}

\maketitle
%\pacs{}
%\keywords{}
%\new

%\setcounter{secnumdepth}{1}

\section{I. Introduction}
The ability to realize photon-photon interactions at the single-photon level will allow for quantum control of optical fields, which has important applications in quantum information science. Photons interact via a nonlinear optical polarization in a material, such as a gas of atoms, but typical light-matter interaction strengths tend to be small so that it is difficult to obtain single-photon nonlinearities. Two-level atoms are predicted to realize single-photon nonlinearities when the optical fields are focused to the size of a wavelength~\cite{Keyes}. However, it is experimentally difficult to focus optical fields to this size, and the ultimate single-photon limit has not yet been reached in systems of two-level atoms.

In order to reach the single-photon limit, researchers have been searching for ways to further enhance the light-matter interaction strength. Promising techniques include placing atoms inside optical cavities or hollow fibers~\cite{PhysRevLett.91.203001,PhysRevLett.102.203902}, employing electromagnetically induced transparency (EIT)~\cite{Nature438}, and using Rydberg blockade~\cite{Dudin18052012,PhysRevLett.109.233602}. Single-photon effects have been observed recently~\cite{Nature436,Tanji-Suzuki02092011,Peyronel,PhysRevLett.112.073901}, which represents a significant step towards realizing quantum logic gates and quantum memories. Enhancing the light-atom interaction strength is also of interest for classical applications in low-light-level nonlinear optics, such as slow light~\cite{Harris} and reducing the threshold for all-optical switching using transverse optical patterns~\cite{Dawes}.

In this paper, we describe a different approach for enhancing the interaction strength that relies on spatial bunching of ultracold two-level atoms in free space. Specifically, we show that spatial organization of atoms in a one-dimensional optical lattice produces a nonlinear susceptibility that is more than two orders of magnitude larger than that attainable via the saturable nonlinearity alone for typical experimental parameters. We thus conclude that single-photon nonlinearities are experimentally feasible in two-level atoms that are spatially organized in an optical lattice.

While it is known that spatial organization of atoms in an optical lattice enhances the light-atom interaction strength~\cite{Ritsch}, existing theoretical models are either restricted to a specific atomic temperature range~\cite{Deutsch, asboth2008, asboth2007, Elsasser} or work in the far-detuned regime where the back-action of the atoms on the lattice-forming optical fields is insignificant and therefore ignored~\cite{asboth2005, Petrosyan, Wu, Nunn, Schilke}. We develop a theoretical model that is valid for all atomic temperatures and accounts for the back-action of the atoms on the lattice-forming optical fields. We are interested in the regime of strong back-action, where small changes in the effective susceptibility of the atomic sample give rise to new physical effects such as transverse optical instabilities~\cite{SchilkeExpTransverse,GreenbergOptExp,Labeyrie} and Bragg scattering~\cite{PhysRevLett.75.2823,PhysRevLett.75.4583,Schilke}.

Our model explicitly connects the results of the zero-temperature models of the optomechanical physics community~\cite{Deutsch, asboth2008} with the finite-temperature models of the nonlinear optics community~\cite{Muradyan, Wang, GreenbergPRA}. The results of our model also provide insight into the effects of high-order nonlinearities. We show that it is important to consider nonlinear optical susceptibilities beyond the third-order response when there is substantial atomic bunching, even at low optical field intensities. As we show, these high-order nonlinear terms enhance (weaken) the effective susceptibility when atoms are tightly bunched in a red (blue) optical lattice\textemdash a result that is supported by multiple experiments (\textit{e.g.},~\cite{Gattobigio, Arnold, GreenbergPRA, Deng}).

We study light-atom interactions in three parameter regimes, depicted in Fig.~\ref{Fig1}, which are delineated by the ratio of the dipole potential energy to the thermal energy of the atoms. In Regime I, the thermal energy exceeds the dipole potential energy, and there is weak or no atomic bunching. In Regime II, the dipole potential energy is on the order of the thermal energy of the atoms, and the majority of the atoms are trapped in the optical lattice. In Regime III, the dipole potential energy greatly exceeds the thermal energy, which results in strong atomic bunching. Figure~\ref{Fig1} also emphasizes that the wavevector of the applied optical fields inside the atomic medium $k^\prime$ can be different than the vacuum wavevector $k$ because of the back-action of the atoms on the optical fields~\cite{Deutsch}. By accounting for this back-action, our model is self-consistent.

This paper is organized as follows. In Sec. II, we present an overview of our model. In Sec. III, we calculate the normalized density distribution and the material susceptibility for the specific case of equal-intensity, frequency-degenerate counterpropagating fields. In Sec. IV, we analyze the results of our model in Regimes I, II, and III, and we discuss the impact on the light-atom interaction from both the nonlinearity that arises due to atomic bunching and the saturable nonlinearity, which couples the optical fields to the internal states of the atoms. In Sec. V, we summarize the new insights our model provides into the enhanced light-atom interaction strengths achievable via atomic bunching of ultracold atoms in an optical lattice.

\section{II. Theoretical Model}
Two-level atoms interact with optical fields according to the effective susceptibility
\begin{equation}
\chi_{\text{eff}}(z)=\frac{-6\pi}{k_{eg}^3}\eta(z)\frac{2\tilde{\Delta}-i}{1+4\tilde{\Delta}^2}\frac{1}{1+I(z)/I_{s\tilde{\Delta}}},
\label{chieff}
\end{equation}
which is the fundamental quantity that describes the light-atom interaction strength~\cite{Boyd}. Here, $\eta(z)$ is the density distribution of atoms, $\tilde{\Delta}=\Delta/\Gamma$, $\Delta=\omega-\omega_{eg}$ is the detuning, defined as the frequency difference between the vacuum applied field frequency $\omega$ and the atomic resonant frequency $\omega_{eg}$, $\Gamma$ is the natural linewidth of the atomic transition, and $k_{eg}=\omega_{eg}/c$. We take $I(z)=2\epsilon_0c\big<\vec{E}(z,t)\cdot\vec{E}^*(z,t)\big>_t$ as the total optical field intensity, where $\epsilon_0$ is the permittivity of free space, $c$ is the speed of light in vacuum, and $\left<\right>_t$ denotes a time average. We also define $I_{s\tilde{\Delta}}=I_s(1+4\tilde{\Delta}^2)$ as the off-resonant saturation intensity, where $I_s=4\epsilon_0c\hbar^2\Gamma^2/|\vec{\mu}|^2$ is the resonant saturation intensity, and $\vec{\mu}$ is the dipole moment. The factor $(1+I(z)/I_{s\tilde{\Delta}})^{-1}$ in Eq.~\ref{chieff} corresponds to the saturable nonlinearity. For the purposes of this paper, we consider detunings that are large enough so that $\chi_{\text{eff}}(z)$ is essentially real, and hence we neglect absorption. Also, we only consider the case where the atoms are in steady-state and thermal equilibrium, and where they do not experience a net radiation pressure force.

\begin{figure}
\begin{center}
 \includegraphics[scale=0.16]{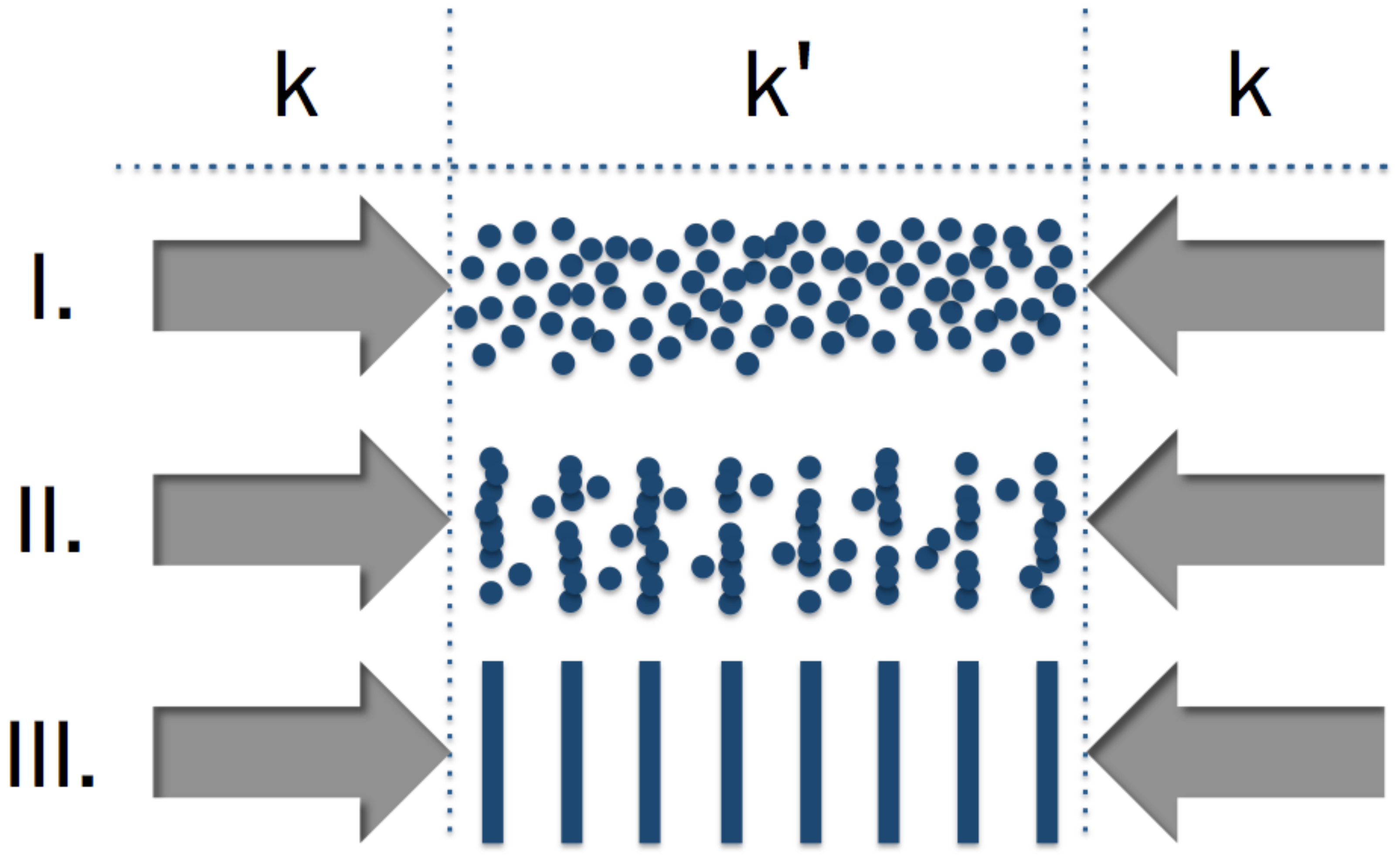}
 \caption{Counterpropagating optical fields (vacuum wavenumber $k$ and wavenumber in medium $k^\prime$) are applied to a gas of atoms. Three regimes are depicted: I. Homogeneous; II. Weak bunching; III. Tight bunching.}
 \label{Fig1}
  \end{center}
 \end{figure}

It is useful to analyze a limiting case of Eq.~\ref{chieff} before we describe the specifics of our model. We take the density distribution $\eta(z)$ to be the steady-state solution of the Fokker-Planck equation, given by
\begin{equation}
\eta(z)=n_a\tilde{\eta}\text{exp}[-U(z)/k_BT],
\label{densitygen}
\end{equation}
where $n_a$ is the average atomic density, $\tilde{\eta}$ is a normalization constant, $U(z)$ is the dipole potential, $k_B$ is Boltzmann's constant, and $T$ is the atomic temperature. The dipole potential for a two-level atom and moderately large detunings (\textit{e.g.}, $|\tilde{\Delta}|\gtrsim3$ for $\text{max}\left[I(z)\right]/I_s<1$) is just the AC Stark Shift. Therefore,
\begin{equation}
\frac{U(z)}{k_BT}=\frac{\tilde{\Delta}I(z)}{I_{s\tilde{\Delta}}\tilde{T}},
\label{dipoledef}
\end{equation}
where $\tilde{T}=T/T_D$ is the temperature normalized by the Doppler temperature $T_D=\hbar\Gamma/2k_B$. If we consider the limiting case $\text{max}[U(z)]/k_BT\ll1$ and $\text{max}[I(z)]/I_{s\tilde{\Delta}}\ll1$, Eq.~\ref{chieff} is given approximately by
\begin{equation}
\chi_{\text{eff}}(z)\approx\chi_{\text{lin}}\left[1-\left(\frac{\tilde{\Delta}}{\tilde{T}}\frac{I(z)-\left<I(z)\right>}{I_{s\tilde{\Delta}}}+\frac{I(z)}{I_{s\tilde{\Delta}}}\right)\right],
\label{chieffapprox}
\end{equation}
where the factor $\chi_{\text{lin}}=-6\pi(2\tilde{\Delta})n_a/[k_{eg}^3(1+4\tilde{\Delta}^2)]$ is the linear susceptibility, and $\left<I(z)\right>$ denotes the spatially averaged value of the intensity. The second term in Eq.~\ref{chieffapprox} corresponds to the bunching-induced nonlinearity, and the third term corresponds to the saturable (Kerr) nonlinearity. As we show in Sec. III, the bunching-induced nonlinearity arises only due to the spatially dependent part of the intensity. For many applications, such as wave-mixing, it is the spatially dependent part of $\chi_{\text{eff}}(z)$ that is of interest. Even though both the bunching-induced nonlinearity and the saturable nonlinearity scale with $I(z)/I_{s\tilde{\Delta}}$, the bunching-induced nonlinearity may be much larger than the saturable nonlinearity for a gas cooled to sub-Doppler temperatures. Below, we demonstrate the impact of this bunching-induced effect on $\chi_{\text{eff}}$ over all atomic temperature regimes---not limited to the approximations that were used to obtain Eq.~\ref{chieffapprox}.

We consider a one-dimensional optical lattice created by two frequency-degenerate, counterpropagating optical fields incident on a sample of two-level atoms, as depicted in Fig.~\ref{Fig1}. The total applied electric field is then
\begin{equation}
\vec{E}(z,t)=\vec{F}(z,t)e^{i(kz-\omega t)}+\vec{B}(z,t)e^{i(-kz-\omega t)}+\text{c.c.},
\label{Efield}
\end{equation}
where $k$ is the wavenumber of the optical fields in vacuum (mnemonics $F$ for forward and $B$ for backward).

Because atoms spatially organize into the potential minima of an optical lattice, the periodicity of the density distribution equals the periodicity of the intensity distribution inside the atomic medium~\cite{Deutsch, asboth2008}. It is therefore convenient to define the density distribution $\eta(z)$ via the Floquet expansion
\begin{equation}
\eta(z)=n_a\sum_{j=-\infty}^{\infty}\tilde{\eta}_j(z)e^{j2ikz}.
\label{density}
\end{equation}
The coefficients $\tilde{\eta}_j(z)$ are derived in Sec. III and have a slowly varying position dependence that accounts for the back-action of the atoms on the optical fields. Contrary to our approach, Refs.~\cite{asboth2005, Petrosyan, Schilke} take the periodicity of the density distribution to be equal to that of the vacuum intensity distribution, which is a very good approximation when the optical fields are far-detuned and the back-action is small. However, we are interested in the regime of strong back-action where one can achieve strong nonlinear effects at low light levels.

We investigate the interaction between the applied optical fields and the atomic medium via the polarization $\vec{P}=\epsilon_0\chi_{\text{eff}}(z)\vec{E}$ using Eqs.~\ref{chieff} and \ref{density}~\cite{Boyd}. We consider the regime where $\text{max}\left[I(z)\right]/\tilde{I}_{s\tilde{\Delta}}\ll1$, so that $(1+I(z)/I_{s\tilde{\Delta}})^{-1}\approx(1-I(z)/I_{s\tilde{\Delta}})$. For $|\tilde{\Delta}|\gtrsim3$, this requires only that $\text{max}\left[I(z)\right]/I_s\lesssim1$. Making the rotating wave approximation and considering both steady-state field amplitudes and parallel optical field polarizations $\big(\vec{F}(z)\big|\big|\vec{B}(z)\big)$, the wave equation gives rise to the coupled amplitude equations
\begin{widetext}
\begin{multline}
\frac{\partial F}{\partial z}=\frac{ik}{2}\chi_{\text{lin}}\Bigg\{\left[\tilde{\eta}_0(z)-\frac{4\epsilon_0c}{I_{s\tilde{\Delta}}}\left(\left[|F|^2+|B|^2\right]\tilde{\eta}_0(z)+FB^*\tilde{\eta}_{-1}(z)+F^*B\tilde{\eta}_{1}(z)\right)\right]F+\\
\left[\tilde{\eta}_1(z)e^{2ikz}-\frac{4\epsilon_0c}{I_{s\tilde{\Delta}}}\left(\left[|F|^2+|B|^2\right]\tilde{\eta}_1(z)e^{2ikz}+FB^*e^{2ikz}\tilde{\eta}_{0}(z)+F^*Be^{-2ikz}\tilde{\eta}_{2}(z)e^{4ikz}\right)\right]Be^{-2ikz}\Bigg\}
\label{F1}
\end{multline}
and
\begin{multline}
\frac{\partial B}{\partial z}=-\frac{ik}{2}\chi_{\text{lin}}\Bigg\{\left[\tilde{\eta}_0(z)-\frac{4\epsilon_0c}{I_{s\tilde{\Delta}}}\left(\left[|F|^2+|B|^2\right]\tilde{\eta}_0(z)+FB^*\tilde{\eta}_{-1}(z)+F^*B\tilde{\eta}_{1}(z)\right)\right]B+\\
\left[\tilde{\eta}_{-1}(z)e^{-2ikz}-\frac{4\epsilon_0c}{I_{s\tilde{\Delta}}}\left(\left[|F|^2+|B|^2\right]\tilde{\eta}_{-1}(z)e^{-2ikz}+FB^*e^{2ikz}\tilde{\eta}_{-2}(z)e^{-4ikz}+F^*Be^{-2ikz}\tilde{\eta}_{0}(z)\right)\right]Fe^{2ikz}\Bigg\},
\label{B1}
\end{multline}
\end{widetext}
where $F\equiv F(z)$, $B\equiv B(z)$, and we have only retained terms that have equal or nearly equal spatial variations. In Eqs.~\ref{F1} and \ref{B1}, the first term in square brackets on the right-hand-side gives rise to the dispersion of the optical fields as they propagate through the atomic medium, and the other represents the nonlinear coupling between the forward and backward fields. In the remainder of this paper, we treat the special case of equal-intensity counterpropagating fields, which suppresses the radiation pressure force and allows us to investigate solely the effects of the dipole potential that gives rise to atomic bunching.

\section{III. Uniform Optical Lattice}
Under our conditions, we find that each optical field experiences the same susceptibility, and hence each optical field has a wavevector $k^\prime=nk$ inside the atomic medium, where the index of refraction $n\simeq1+\chi_{\text{eff}}/2$~\cite{Boyd}. Here, $\chi_{\text{eff}}$ is independent of $z$ and contains only those terms from Eq.~\ref{chieff} that are spatially matched to each optical field, \textit{i.e.}, only those terms that are kept in Eqs.~\ref{F1} and \ref{B1}. We can therefore take the optical field amplitudes to have the forms
\begin{equation}
F(z)=\tilde{F}e^{ik(\chi_{\text{eff}}/2)z}\text{  and  }B(z)=\tilde{B}e^{-ik(\chi_{\text{eff}}/2)z},
\end{equation}
where $\tilde{F}$ and $\tilde{B}$ are independent of $z$. Analogously, we define
\begin{equation}
\eta_j(z)=n_a\tilde{\eta}_je^{j2ik(\chi_{\text{eff}}/2)z},
\end{equation}
where the coefficients $\tilde{\eta}_j$ are independent of $z$ because the periodicity of the density distribution in Eq.~\ref{density} exactly equals the intensity distribution inside the medium. Imposing the boundary conditions $F(-L/2)=B(L/2)$ for a medium of length $L$, Eqs. \ref{F1} and \ref{B1} allow us to find the susceptibility
\begin{equation}
\chi_{\text{eff}}=\chi_{\text{lin}}\left[\tilde{\eta}_0+\tilde{\eta}_{\pm1}-\frac{\tilde{I}_{\tilde{\Delta}}}{2}\left(3\tilde{\eta}_0+\tilde{\eta}_{\mp1}+3\tilde{\eta}_{\pm1}+\tilde{\eta}_{\pm2}\right)\right],
\label{delta2}
\end{equation}
where $\tilde{I}_{\tilde{\Delta}}=\left<I(z)\right>/I_{s\tilde{\Delta}}$ defines the spatially averaged value of the total intensity due to both optical fields normalized by the off-resonant saturation intensity, and
\begin{equation}
\tilde{\eta}_j=\frac{1}{\lambda^\prime/2}\int_{-\lambda^\prime/4}^{\lambda^\prime/4}\frac{\eta(z)}{n_a}e^{-j2ik^\prime z}\text{d}z
\end{equation}
with $\lambda^\prime=2\pi/k^\prime$.

We calculate the density distribution from Eq.~\ref{densitygen}, where the position-independent part of $U(z)$ does not contribute to the dipole force and thus does not give rise to atomic bunching. We absorb the position-independent part of $U(z)$ into a new normalization constant $\tilde{\eta}^\prime$, and the Fourier coefficients become
\begin{equation}
\tilde{\eta}_j=\frac{\tilde{\eta}^\prime k^\prime}{\pi}\int_{-\pi/2k^\prime}^{\pi/2k^\prime}\text{exp}\left[-\tilde{U}_{\tilde{T}}\text{cos}(2k^\prime z)\right]e^{-j2ik^\prime z}\text{d}z,
\label{etaj}
\end{equation}
where
\begin{equation}
\tilde{U}_{\tilde{T}}=\frac{\tilde{\Delta}\tilde{I}_{\tilde{\Delta}}}{\tilde{T}}.
\end{equation}

From Eq.~\ref{etaj}, we find that $\tilde{\eta}_j=\tilde{\eta}^\prime I_j(-\tilde{U}_{\tilde{T}})$, where $I_j$ are modified Bessel functions of the first kind of order $j$. We also determine $\tilde{\eta}^\prime$ by integrating the density distribution over one period, \textit{i.e.},
\begin{equation}
\frac{\lambda^\prime}{2}=\tilde{\eta}^\prime\int_{-\pi/2k^\prime}^{\pi/2k^\prime}\text{exp}\left[-\tilde{U}_{\tilde{T}}\text{cos}(2k^\prime z)\right]\text{d}z.
\end{equation}
This results in the relation $\lambda^\prime/2=\pi\tilde{\eta}^\prime I_0(-\tilde{U}_{\tilde{T}})/k^\prime$. Therefore, the normalization constant is
\begin{equation}
\tilde{\eta}^\prime=\frac{1}{I_0(-\tilde{U}_{\tilde{T}})}.
\label{densitynormalization}
\end{equation}
Previous research has indicated the importance of the normalization constant for defining a physically accurate density distribution, but they do not explicitly calculate it and instead either take it to be fixed by experimental conditions~\cite{Muradyan, Wang, Schilke} or do not require it for their analysis~\cite{Tesio}. By explicitly accounting for this normalization constant, our model is applicable to all regimes of atomic bunching.

\begin{figure}[b]
\begin{center}
 \includegraphics[scale=.19]{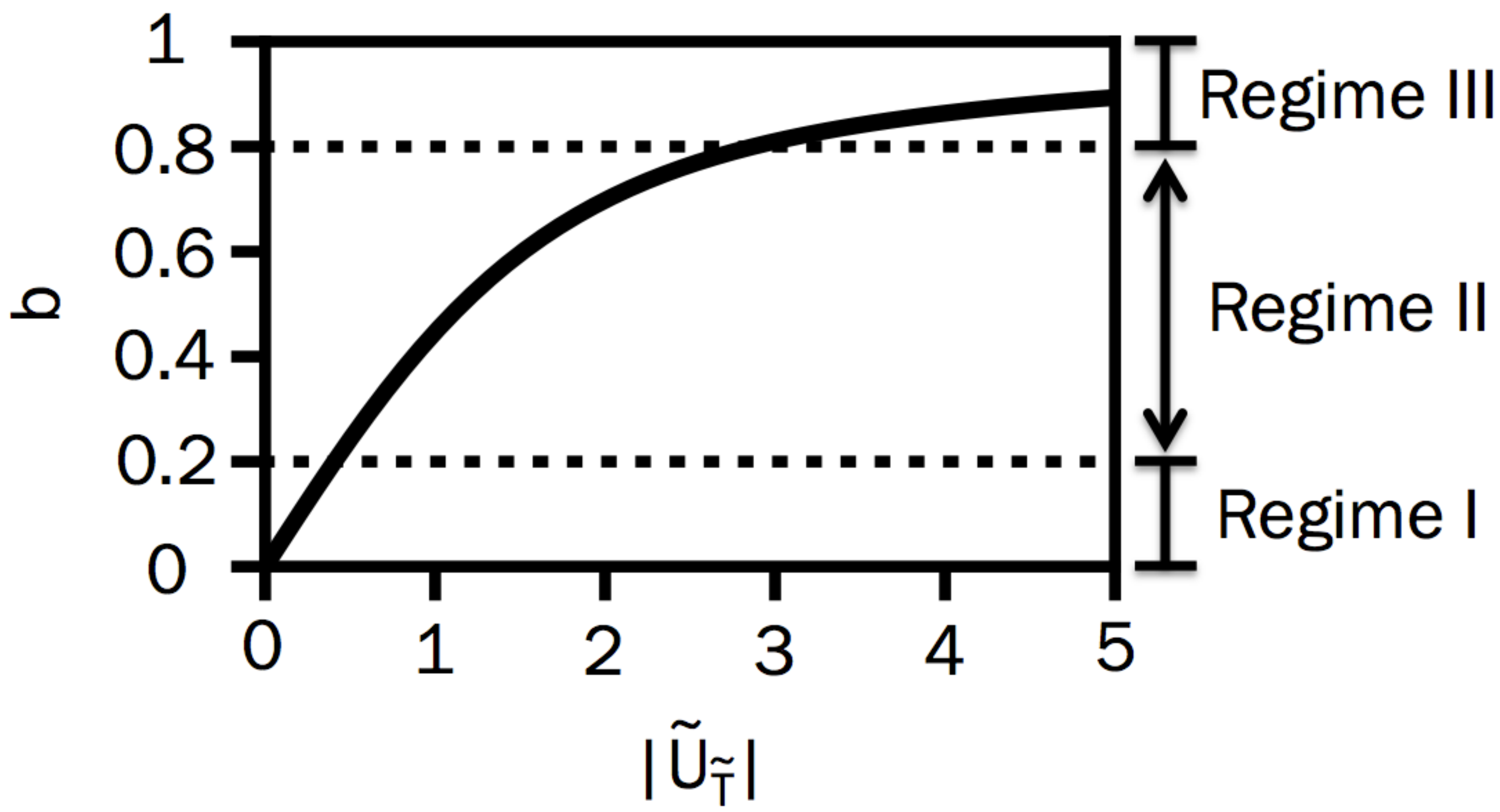}
 \caption{Bunching parameter as a function of the depth of the dipole potential wells of the optical lattice.}
 \label{bunchingparameterplot}
  \end{center}
 \end{figure}

Combining Eqs.~\ref{etaj} and ~\ref{densitynormalization}, the Fourier coefficients are
\begin{equation}
\tilde{\eta}_j=\frac{I_j(-\tilde{U}_{\tilde{T}})}{I_0(-\tilde{U}_{\tilde{T}})}.
\label{FCs}
\end{equation}
We also define the bunching parameter
\begin{equation}
b=|\tilde{\eta}_1|,
\end{equation}
which describes the degree of atomic bunching and distinguishes the three regimes depicted in Fig.~\ref{Fig1}. This is analogous to the bunching parameter introduced in Ref.~\cite{PhysRevA.50.1716}, but where we have only retained the first-order Fourier component of the density distribution because it is the only component that directly contributes to coupling the forward and backward waves. The bunching parameter $b\in[0,1]$, where $b=0$ corresponds to a homogeneous gas and $b=1$ corresponds to the case where the density distribution consists of infinitesimally thin sheets of atoms. Figure~\ref{bunchingparameterplot} shows $b$ as a function of $|\tilde{U}_{\tilde{T}}|$. Here, we define the regimes of atomic bunching using reasonable but arbitrary cutoffs of the bunching parameter. Regime I occurs where $b<0.2$ ($|\tilde{U}_{\tilde{T}}|<0.4$). Regime II occurs where $0.2\le b\le0.8$ ($0.4\le|\tilde{U}_{\tilde{T}}|<2.9$), where the thermal energy of the atoms is on the order of the dipole potential energy. Regime III corresponds to strong spatial localization of the atoms, where $b>0.8$ ($|\tilde{U}_{\tilde{T}}|>2.9$), which is attainable using typical conditions in a magneto-optical trap~\cite{PhysRevLett.75.2823}.

\begin{figure}
\begin{center}
 \includegraphics[scale=.26]{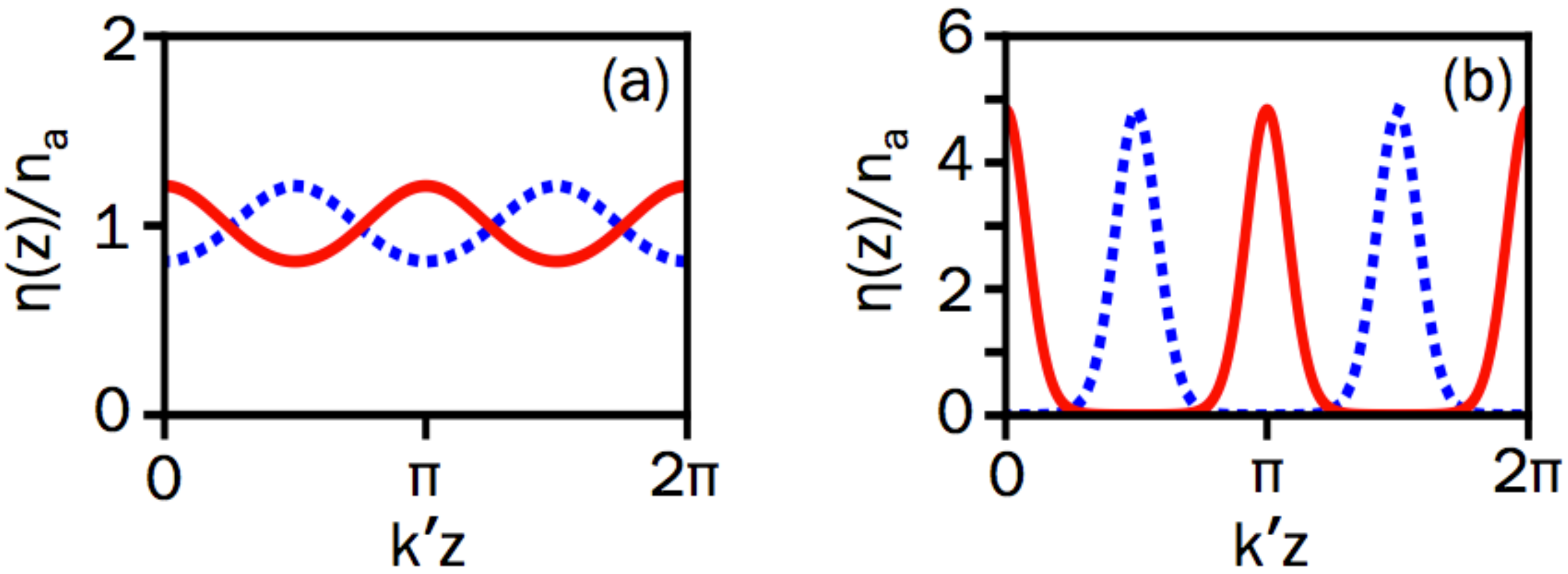}
 \caption{Atomic density distributions for (a) $b=0.01$ $\left(|\tilde{U}_{\tilde{T}}|=0.2\right)$ and (b) $b=0.86$ $\left(|\tilde{U}_{\tilde{T}}|=4\right)$, with $\tilde{\Delta}<0$ (red, solid) and $\tilde{\Delta}>0$ (blue, dashed).}
 \label{densityplots}
  \end{center}
 \end{figure}

Examples of the density distribution are plotted in Fig. \ref{densityplots} for two different bunching parameters. Figure \ref{densityplots}(a) shows Regime I, where the density distribution is weakly modulated about the average atomic density $\eta(z)/n_a=1$. Figure \ref{densityplots}(b) shows the density distribution for Regime III, where the local density greatly exceeds the average density, and the atoms are well-localized to the potential wells. In fact, it is in this parameter regime where the first Bragg scattering experiments for atoms in optical lattices were performed: $b\simeq0.7$ ($|\tilde{U}_{\tilde{T}}|\simeq 2$) in Ref.~\cite{PhysRevLett.75.4583} and $b\simeq0.93$ ($|\tilde{U}_{\tilde{T}}|\simeq 7$) in Ref.~\cite{PhysRevLett.75.2823}). Figure~\ref{densityplots} also shows that, when using red ($\tilde{\Delta}<0$) versus blue ($\tilde{\Delta}>0$) detuned optical fields, the density maxima occur at different locations (phase-shifted by $z=\lambda^\prime/4$). This is consistent with the spatial organization of the atoms into the dipole potential minima, which corresponds to the intensity maxima (minima) for red (blue) optical lattices. 

Combining Eqs.~\ref{delta2} and \ref{FCs}, the effective susceptibility experienced by the optical fields is
\begin{multline}
\chi_{\text{eff}}=\chi_{\text{lin}}\Big[1+\frac{I_1(-\tilde{U}_{\tilde{T}})}{I_0(-\tilde{U}_{\tilde{T}})}+\\
 -\frac{\tilde{I}_{\tilde{\Delta}}}{2}\left(3+4\frac{I_1(-\tilde{U}_{\tilde{T}})}{I_0(-\tilde{U}_{\tilde{T}})}+\frac{I_2(-\tilde{U}_{\tilde{T}})}{I_0(-\tilde{U}_{\tilde{T}})}\right)\Big].
\label{chieffprime}
\end{multline}
Equation~\ref{chieffprime} provides the basis for the analysis in the remainder of this paper, where we study how $\chi_{\text{eff}}$ varies with $\tilde{U}_{\tilde{T}}$ in the three regimes depicted in Fig.~\ref{Fig1}.

In Regime I, a good approximation of $\chi_{\text{eff}}$ is a Taylor expansion to first order in $\tilde{I}_{\tilde{\Delta}}$, which is given by
\begin{equation}
\frac{\chi_{\text{eff}}}{|\chi_{\text{lin}}|}\simeq-\frac{\tilde{\Delta}}{|\tilde{\Delta}|}\left[1-\frac{\tilde{\Delta}\tilde{I}_{\tilde{\Delta}}}{2\tilde{T}}-\frac{3}{2}\tilde{I}_{\tilde{\Delta}}\right]
\hspace{1.5mm}
\text{for}
\hspace{2mm}
b<0.2.
 \label{chiefftaylor}
\end{equation}
This equation is consistent with Eq.~\ref{chieffapprox}, where the factor of $1/2$ appearing in the nonlinear terms in Eq.~\ref{chiefftaylor} arises because only one term in the exponential form of the intensity distribution gives rise to the spatially matched, nonlinear coupling in the wave equation.

In Regime III, a good approximation of $\chi_{\text{eff}}$ is an asymptotic expansion, which is given by
\begin{equation}
\frac{\chi_{\text{eff}}}{|\chi_{\text{lin}}|}\simeq 
\begin{cases}
    2-\frac{\tilde{T}}{2|\tilde{\Delta}|\tilde{I}_{\tilde{\Delta}}}-4\tilde{I}_{\tilde{\Delta}}+\frac{2\tilde{T}}{|\tilde{\Delta}|}& \text{if } \tilde{\Delta}<0\\
    -\frac{\tilde{T}}{2|\tilde{\Delta}|\tilde{I}_{\tilde{\Delta}}}-\frac{3\tilde{T}}{4|\tilde{\Delta}|}& \text{if } \tilde{\Delta}>0.
\end{cases}
\label{asymptoticchiefftaylor}
\end{equation}

\begin{figure}
\begin{center}
\begin{minipage}{.5\textwidth}
 \includegraphics[scale=0.27]{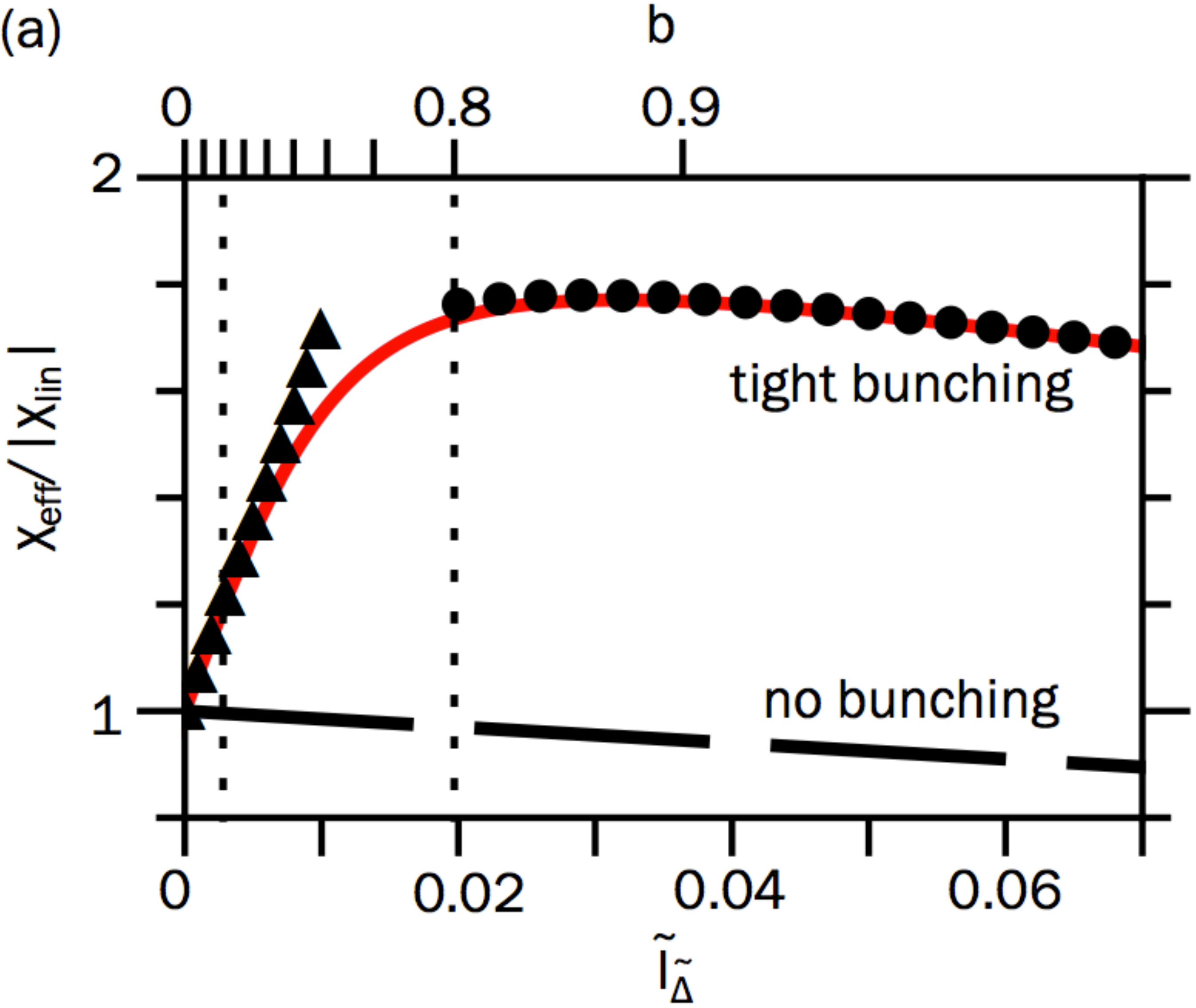}
 \end{minipage}\\
\vspace{3mm}
\begin{minipage}{.5\textwidth}
 \includegraphics[scale=0.27]{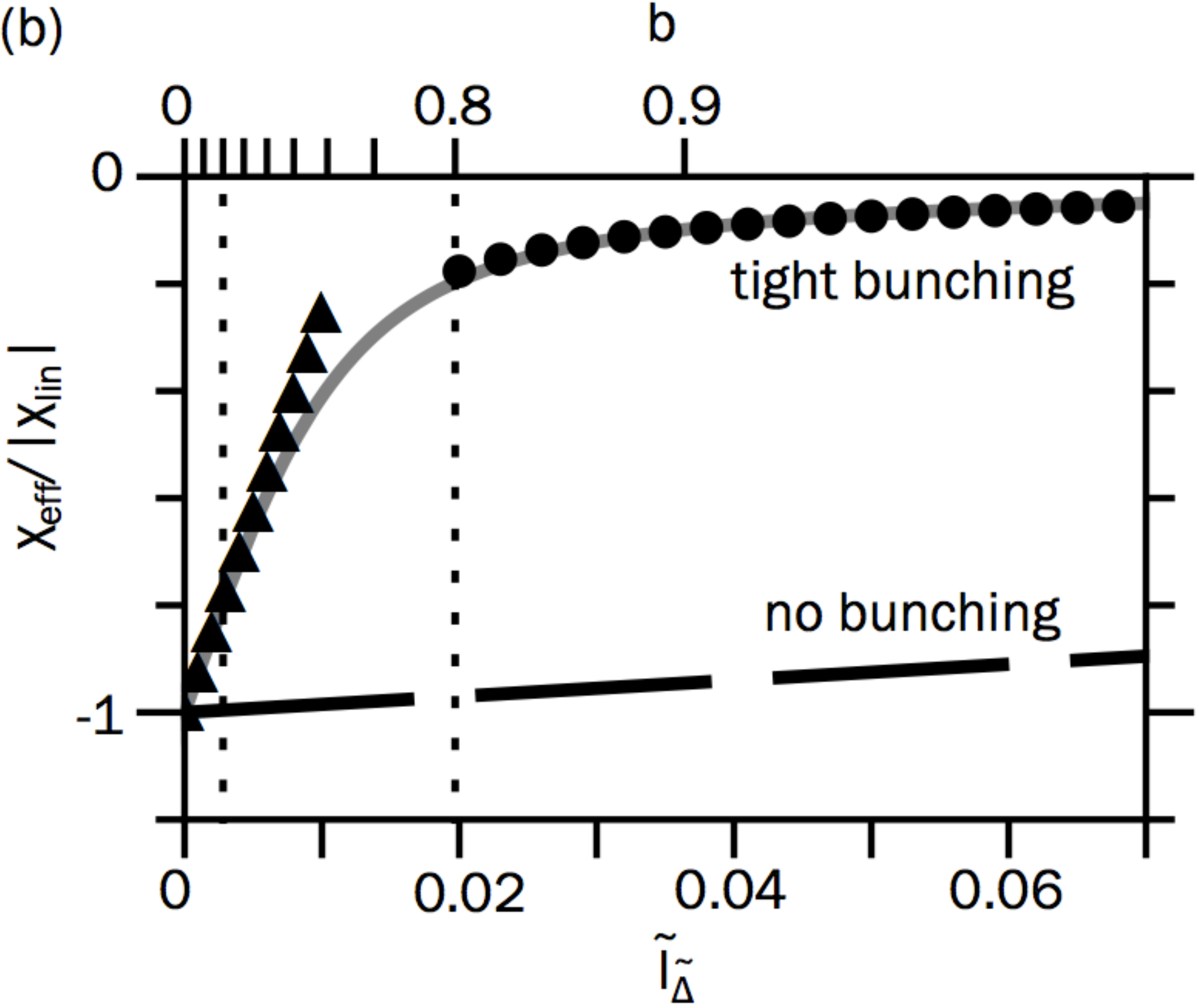}
 \end{minipage}
\caption{The effective susceptibility as functions of $\tilde{I}_{\tilde{\Delta}}$ and $b$ for (a) red detunings ($\tilde{\Delta}=-3$) and (b) blue detunings ($\tilde{\Delta}=3$). The solid curve is the case $\tilde{T}=3/146$, corresponding to $T=3$ $\mu$K and $T_D=146$ $\mu$K for rubidium. The long, dashed line is the case $\tilde{T}\rightarrow\infty$. The triangles represent the Taylor series expansion (Eq.~\ref{chiefftaylor}), and the circles represent the asymptotic expansion (Eq.~\ref{asymptoticchiefftaylor}). The vertical, dashed lines correspond to $b=0.2$ and $b=0.8$, which designate the boundaries between Regimes I, II, and III for increasing intensity}
 \label{ChiZoom}
  \end{center}
 \end{figure}

The typical behavior of $\chi_{\text{eff}}$ is plotted in Fig.~\ref{ChiZoom} as a function of both $\tilde{I}_{\tilde{\Delta}}$ and $b$ along with both the Taylor and asymptotic expansions. The overall scale of each curve in Fig.~\ref{ChiZoom} is set by $\chi_{\text{lin}}$, and the slope of each curve is directly related to the third-order nonlinear optical susceptibility $\chi^{(3)}$. From Eq.~\ref{chiefftaylor}, the third-order nonlinear susceptibility in Regime I is given by
\begin{equation}
\chi^{(3)}\simeq\frac{\tilde{\Delta}}{|\tilde{\Delta}|}\frac{2\epsilon_0c|\chi_{\text{lin}}|}{I_{s\tilde{\Delta}}}\left[\frac{\tilde{\Delta}}{2\tilde{T}}+\frac{3}{2}\right]
\hspace{1.5mm}
\text{for}
\hspace{2mm}
b<0.2,
\label{dchiefftaylor}
\end{equation}
where the first term is the contribution from the bunching-induced nonlinearity and the second is from the saturable (Kerr) nonlinearity.

In the case of a homogeneous gas ($b=0$), $\chi^{(3)}=3\epsilon_0c|\chi_{\text{lin}}|\tilde{\Delta}/I_{s\tilde{\Delta}}|\tilde{\Delta}|$, corresponding to a self-defocusing (self-focusing) nonlinearity for red (blue) detunings. This has important implications for nonlinear optical processes in homogeneous atoms, \textit{e.g.}, transverse optical pattern formation, where patterns can only form when $\chi^{(3)}>0$~\cite{Chiao}. Our model is consistent with experiments that find pattern formation occurs only for blue-detuned optical fields when using a homogeneous gas of atoms~\cite{Dawes}.

Despite the fact that homogeneous atomic samples give rise to strong nonlinear susceptibilities~\cite{Schmidt1996,Dawes}, the nonlinear susceptibility can be further enhanced by using sub-Doppler-cooled atoms, which is indicated by the inverse dependence of $\chi^{(3)}$ on $\tilde{T}$ in Eq.~\ref{dchiefftaylor}. This appears as the steep slopes in Fig.~\ref{ChiZoom} for $b<0.2$ and $\tilde{T}=3/146$. For a gas of rubidium atoms at $\tilde{T}=3$~$\mu$K\textemdash achievable using Sisyphus cooling, for example~\cite{GreenbergPRA, GreenbergOptExp}\textemdash and $|\tilde{\Delta}|=7$, $\chi^{(3)}$ is more than two orders of magnitude larger than in the homogeneous case.

In the homogeneous case, the only contribution to the nonlinear susceptibility is the saturable nonlinearity, which itself is predicted to reach the single-photon nonlinearity threshold when the optical fields are focused to the size of a wavelength~\cite{Keyes}. With the enhanced material response due to atomic bunching, we predict that one can achieve single-photon nonlinearities, \textit{e.g.} single-photon optical switching using transverse optical pattern formation~\cite{Dawes}, without the requirement of focusing the optical fields to their ultimate limit. It is this regime that is of interest for low-light-level nonlinear optical applications and the search for photon-photon interactions at the single-photon level.

In the case of strong atomic bunching ($b>0.8$), $\chi^{(3)}$ is only greater than the case of a homogeneous gas for red detunings. From Eq.~\ref{asymptoticchiefftaylor}, $\chi^{(3)}$ in Regime III is 
\begin{equation}
\chi^{(3)}\simeq
\begin{cases}
    \frac{2\epsilon_0c|\chi_{\text{lin}}|}{I_{s\tilde{\Delta}}}\left(\frac{\tilde{T}}{2|\tilde{\Delta}|\tilde{I}_{\tilde{\Delta}}^2}-4\right)& \text{if } \tilde{\Delta}<0\\
    \frac{2\epsilon_0c|\chi_{\text{lin}}|}{I_{s\tilde{\Delta}}}\frac{\tilde{T}}{2|\tilde{\Delta}|\tilde{I}_{\tilde{\Delta}}^2}& \text{if } \tilde{\Delta}>0.
\end{cases}
\label{dasymptoticchiefftaylor}
\end{equation}
Here, $\chi^{(3)}$ depends on the intensity because there are high-order nonlinear terms in $\chi_{\text{eff}}$ that are larger than the third-order contribution and cannot be neglected. This will be discussed further in Sec. IV.

Equation~\ref{dasymptoticchiefftaylor} indicates that, for larger intensities, $\chi^{(3)}\rightarrow-8\epsilon_0c|\chi_{\text{lin}}|/I_{s\tilde{\Delta}}$ for red detunings, and $\chi^{(3)}\rightarrow0$ for blue detunings, which are both independent of the atomic temperature. This arises from the fact that the bunching-induced nonlinearity plays a less substantial role once the atoms are confined tightly. In addition, $\chi^{(3)}$ is a factor of $8/3$ larger in magnitude than in the homogeneous case. However, both $\chi_{\text{eff}}$ and $\chi^{(3)}$ approach zero for blue detunings, which indicates that the atoms are not interacting with the optical fields. This is consistent with the fact that tightly bunched atoms interact strongly with the intensity maxima (minima) of a red (blue) optical lattice, and $\chi^{(3)}$ is therefore stronger (weaker) than for the homogeneous gas.

Figure~\ref{ChiZoom}(a) indicates that there is a local maximum in $\chi_{\text{eff}}$ for red detunings and increasing $\tilde{I}_{\tilde{\Delta}}$ in Regime III. From Eq.~\ref{dasymptoticchiefftaylor}, this critical point occurs at $\tilde{I}_{\tilde{\Delta}}\simeq\sqrt{\tilde{T}/8|\tilde{\Delta}|}$, or when
\begin{equation}
|\tilde{U}_{\tilde{T}}|\simeq\sqrt{\frac{|\tilde{\Delta}|}{8\tilde{T}}}.
\label{critint}
\end{equation}
This critical point corresponds to the case where the nonlinearity transitions from self-focusing to self-defocusing for increasing intensities, which, for red detunings, corresponds to the condition at which the saturable nonlinearity begins to dominate over the bunching-induced nonlinearity.

However, there does not exist a critical point for atoms in a blue optical lattice, which spatially bunch into the standing-wave nodes. In this case, increasing the depth of the dipole potential wells only reduces the number of atoms that can interact with the optical fields. This supports multiple experiments that find that nonlinear optical processes occurring in the tight bunching regime are induced at higher intensities using blue-detuned optical fields~\cite{GreenbergPRA, GreenbergOptExp, Gattobigio, Arnold, Deng}.

In order to better understand the trends appearing in Fig.~\ref{ChiZoom} for the different bunching regimes, we will next investigate the different contributions to $\chi_{\text{eff}}$ due to the bunching-induced and saturable nonlinearities.

\section{IV. Interference Between Competing Nonlinearities}
To understand the physical effects that contribute to the susceptibility, we decompose $\chi_{\text{eff}}$ into parts as
\begin{equation}
\chi_{\text{eff}}=\chi_{\text{lin}}+\chi_{\text{bunching}}+\chi_{\text{s}}+\chi_{\text{bunching+s}},
\label{deltasum}
\end{equation}
indicating the contributions due to linear effects, the bunching-induced nonlinearity, the saturable nonlinearity, and the combined effects of these two nonlinearities, respectively. We find that
\begin{equation}
\chi_{\text{bunching}}=\chi_{\text{lin}}\frac{I_1(-\tilde{U}_{\tilde{T}})}{I_0(-\tilde{U}_{\tilde{T}})},
\label{gammabunching}
\end{equation}
\begin{equation}
\chi_{\text{s}}=-\frac{3\chi_{\text{lin}}\tilde{I}_{\tilde{\Delta}}}{2},
\label{gammakerr}
\end{equation}
and
\begin{equation}
\chi_{\text{bunching+s}}=-\frac{\chi_{\text{lin}}\tilde{I}_{\tilde{\Delta}}}{2}\left[4\frac{I_1(-\tilde{U}_{\tilde{T}})}{I_0(-\tilde{U}_{\tilde{T}})}+\frac{I_2(-\tilde{U}_{\tilde{T}})}{I_0(-\tilde{U}_{\tilde{T}})}\right],
\label{gammabunchingkerr}
\end{equation}
where the numerical factor of $3$ in $\chi_{\text{s}}$ appears because the spatially independent part of the intensity polarizes the atoms and contributes to the saturable nonlinearity. This numerical factor is absent in $\chi_{\text{bunching}}$ because only the spatially dependent part of the dipole potential gives rise to atomic bunching.

\begin{figure}
\raisebox{-.5\height}{\includegraphics[scale=.22]{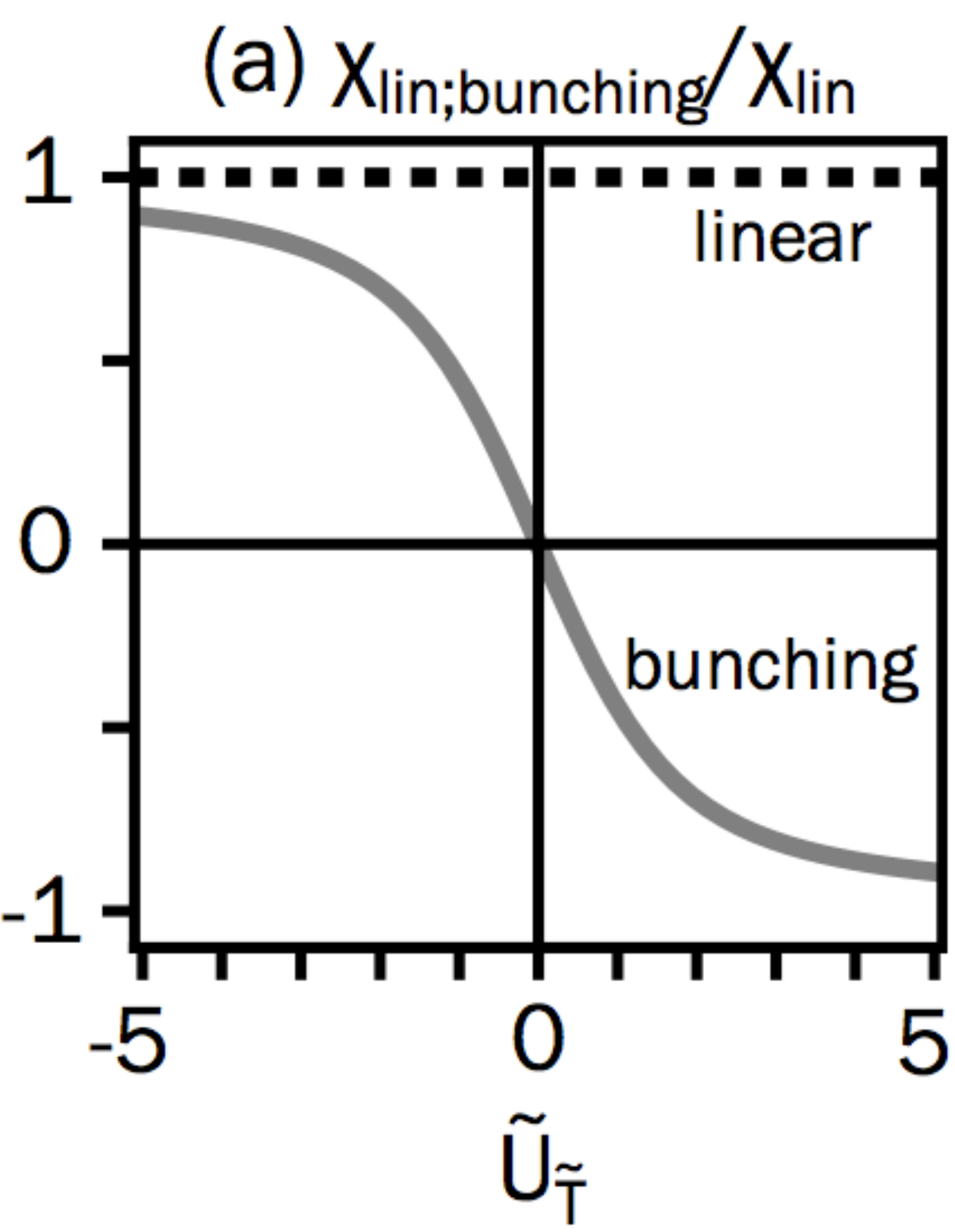}}
\hspace{4mm}
\raisebox{-.5\height}{\includegraphics[scale=.22]{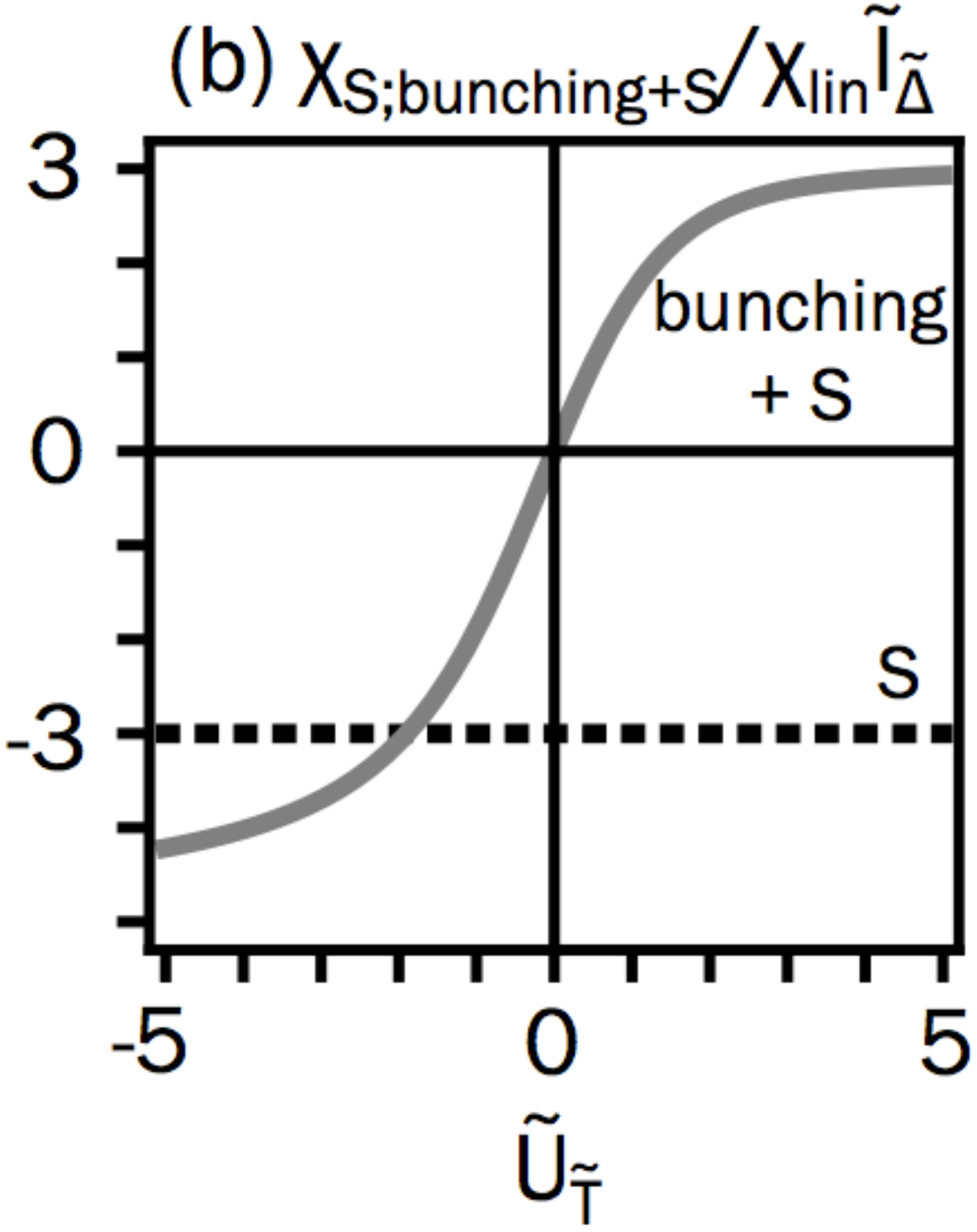}}
 \caption{Components of $\chi_{\text{eff}}$ as functions of $\tilde{U}_{\tilde{T}}$. (a) $\chi_{\text{lin}}/\chi_{\text{lin}}$ (dashed line) and $\chi_{\text{bunching}}/\chi_{\text{lin}}$ (solid curve). (b) $\chi_{\text{s}}/\chi_{\text{lin}}\tilde{I}_{\tilde{\Delta}}$ (dashed line) and $\chi_{\text{bunching+s}}/\chi_{\text{lin}}\tilde{I}_{\tilde{\Delta}}$ (solid curve).}
 \label{wavevectors}
 \end{figure}

The relative contributions of each nonlinear effect in Eqs.~\ref{gammabunching}-\ref{gammabunchingkerr} are illustrated in Fig.~\ref{wavevectors}. If we first consider blue detunings ($\tilde{U}_{\tilde{T}}>0$) in Figs.~\ref{wavevectors}(a) and (b), the contributions $\chi_{\text{lin}}$ and $\chi_{\text{bunching}}$ have opposite signs, as do $\chi_{\text{bunching+s}}$ and $\chi_{\text{s}}$. Therefore, each of these sets of terms interfere destructively, resulting in a very small $\chi_{\text{eff}}$. This is the expected result for atoms localized in the intensity minima and supports the trend in Fig.~\ref{ChiZoom}(b) for larger intensities, where the potential depth is larger.

For red detunings ($\tilde{U}_{\tilde{T}}<0$), Figs.~\ref{wavevectors}(a) and (b) show that $\chi_{\text{lin}}$ and $\chi_{\text{bunching}}$ have the same sign, which is opposite the sign of $\chi_{\text{s}}$ and $\chi_{\text{bunching+s}}$. However, destructively interfering processes in this case do not have identical dependences on $\tilde{I}_{\tilde{\Delta}}$, and the relative strengths of the nonlinear contributions to $\chi_{\text{eff}}$ depend on whether one is below or above the critical point given by Eq.~\ref{critint}.

The case of very tight atomic bunching ($b>0.8$, $|\tilde{U}_{\tilde{T}}|>2.9$) corresponds to well-localized atoms and is the regime relevant to optomechanical-type systems. In the limit $\tilde{T}\rightarrow0$, we expect the results of our model to match the results of the zero-temperature models of Refs.~\cite{Deutsch, asboth2008}, which treat the bunched atoms as infinitely thin dielectric sheets. These works show that the wavevector in the medium is larger than (identical to) the vacuum wavevector for red (blue) optical lattices. Taking $\tilde{T}\rightarrow0$ in Eq. \ref{chieffprime}, the wavevector in the medium $k^\prime=k(1+\chi_{\text{eff}}/2)$ becomes
\begin{equation}
 k^\prime=
  \begin{cases}
   k\left[1+\chi_{\text{lin}}\left(1-2\tilde{I}_{\tilde{\Delta}}\right)\right]
 & \text{if } \tilde{\Delta} \le 0 \\
   k
       & \text{if } \tilde{\Delta} > 0,
  \end{cases}
\label{kprimetightbunching}
\end{equation}
which agrees with the results of Refs.~\cite{Deutsch, asboth2008}.

This result would not have been obtained if we had used approximate expressions for $\chi_{\text{bunching}}$ and $\chi_{\text{bunching+s}}$. Specifically, Eqs.~\ref{gammabunching} and \ref{gammabunchingkerr} contain high-order nonlinear contributions (where $\chi_{\text{bunching+s}}$ is fifth-order in the lowest-order Taylor expansion of $\tilde{U}_{\tilde{T}}$), which are neglected in most nonlinear optical models that consider $\tilde{I}_{\tilde{\Delta}}\ll1$~\cite{Boyd, Muradyan}.

These high-order nonlinear terms become less important as the bunching parameter decreases. In Regime I, the role played by $\chi_{\text{bunching+s}}$ is far less substantial, and $\chi_{\text{eff}}$ for $b<0.2$ is given by Eq.~\ref{chiefftaylor}. We are only in the third-order nonlinear optical regime for $b<0.2$.

Equation~\ref{chiefftaylor} provides insights into predictions for nonlinear optical processes, \textit{e.g.}, transverse optical pattern formation. The threshold condition at which transverse optical pattern formation can occur is approximately $k\chi_{\text{eff}}^{\text{NL}}L\simeq\pi/2$, where $\chi_{\text{eff}}^{\text{NL}}=\chi_{\text{eff}}-\chi_{\text{lin}}$ is the nonlinear part of the effective susceptibility~\cite{Firth90}. From Eq.~\ref{chiefftaylor}, this threshold condition in the third-order nonlinear optical regime is
\begin{equation}
\frac{\left<I(z)\right>}{I_s}\simeq\frac{k^2}{12n_aL\tilde{\Delta}}\frac{\left(1+4\tilde{\Delta}^2\right)^2}{3+\tilde{\Delta}/\tilde{T}}.
\end{equation}
The minimum intensity threshold is therefore obtained by maximizing $n_a$, minimizing $\tilde{\Delta}$, and minimizing $\tilde{T}$.

In addition, pattern formation will only occur in the third-order nonlinear regime when $(\chi_{\text{bunching}}+\chi_{\text{s}})>0$~\cite{Chiao}. Based on Eq.~\ref{chiefftaylor}, this condition is always satisfied for blue detunings $(\chi_{\text{lin}}<0)$, but it is only met for red detunings $(\chi_{\text{lin}}>0)$ when $|\chi_{\text{bunching}}|>|\chi_{\text{s}}|$. For sufficiently cold atoms, $|\chi_{\text{bunching}}|\gg|\chi_{\text{s}}|$, and the threshold for pattern formation will be the same for both detunings. However, when the atomic temperature approaches the Doppler temperature ($\tilde{T}\simeq1$), $|\chi_{\text{bunching}}|$ and $|\chi_{\text{s}}|$ interfere constructively (destructively) for blue (red) optical lattices. This implies that the threshold for transverse pattern formation should occur at lower optical intensities for blue detunings, which agrees with the predictions of Ref.~\cite{Muradyan}. Only on the self-focusing side of the critical point of Eq.~\ref{critint} will transverse pattern formation occur for red detunings.

\section{V. Conclusions}
In conclusion, we show that spatially organized two-level atoms give rise to a large nonlinear material response at very low optical intensities and that the achievable nonlinear susceptibility is more than two orders of magnitude larger for spatially organized two-level atoms than for a homogeneous sample of warm atoms. We predict that, by using sub-Doppler-cooled, two-level atoms, single-photon nonlinear optics is experimentally feasible. Our model uncovers new insights into the competing effects of atomic bunching and the saturable nonlinearity and shows that high-order nonlinear terms play a substantial role even at low intensities for sufficiently cold atoms. Our model is consistent with the results obtained in other theoretical models and experimental findings over a wide range of temperatures and is therefore a general model that may be used to describe a broad scope of low-light-level physical systems. Future extensions of this work will need to consider Sisyphus cooling and multi-wave-mixing processes~\cite{GreenbergEPL}, which will be useful in predicting thresholds for more complicated nonlinear optical processes, $\textit{e.g.}$, transverse optical pattern formation~\cite{GreenbergOptExp}.

\section{Acknowledgements}
We gratefully acknowledge the financial support of the National Science Foundation through Grant PHY-1206040.

\bibliography{SchmittbergerPRAreferences}

\begin{thebibliography}{40}%
\makeatletter
\providecommand \@ifxundefined [1]{%
 \@ifx{#1\undefined}
}%
\providecommand \@ifnum [1]{%
 \ifnum #1\expandafter \@firstoftwo
 \else \expandafter \@secondoftwo
 \fi
}%
\providecommand \@ifx [1]{%
 \ifx #1\expandafter \@firstoftwo
 \else \expandafter \@secondoftwo
 \fi
}%
\providecommand \natexlab [1]{#1}%
\providecommand \enquote  [1]{``#1''}%
\providecommand \bibnamefont  [1]{#1}%
\providecommand \bibfnamefont [1]{#1}%
\providecommand \citenamefont [1]{#1}%
\providecommand \href@noop [0]{\@secondoftwo}%
\providecommand \href [0]{\begingroup \@sanitize@url \@href}%
\providecommand \@href[1]{\@@startlink{#1}\@@href}%
\providecommand \@@href[1]{\endgroup#1\@@endlink}%
\providecommand \@sanitize@url [0]{\catcode `\\12\catcode `\$12\catcode
  `\&12\catcode `\#12\catcode `\^12\catcode `\_12\catcode `\%12\relax}%
\providecommand \@@startlink[1]{}%
\providecommand \@@endlink[0]{}%
\providecommand \url  [0]{\begingroup\@sanitize@url \@url }%
\providecommand \@url [1]{\endgroup\@href {#1}{\urlprefix }}%
\providecommand \urlprefix  [0]{URL }%
\providecommand \Eprint [0]{\href }%
\providecommand \doibase [0]{http://dx.doi.org/}%
\providecommand \selectlanguage [0]{\@gobble}%
\providecommand \bibinfo  [0]{\@secondoftwo}%
\providecommand \bibfield  [0]{\@secondoftwo}%
\providecommand \translation [1]{[#1]}%
\providecommand \BibitemOpen [0]{}%
\providecommand \bibitemStop [0]{}%
\providecommand \bibitemNoStop [0]{.\EOS\space}%
\providecommand \EOS [0]{\spacefactor3000\relax}%
\providecommand \BibitemShut  [1]{\csname bibitem#1\endcsname}%
\let\auto@bib@innerbib\@empty
%</preamble>
\bibitem [{\citenamefont {Keyes}(1970)}]{Keyes}%
  \BibitemOpen
  \bibfield  {author} {\bibinfo {author} {\bibfnamefont {R.~W.}\ \bibnamefont
  {Keyes}},\ }\href@noop {} {\bibfield  {journal} {\bibinfo  {journal}
  {Science}\ }\textbf {\bibinfo {volume} {168}},\ \bibinfo {pages} {796}
  (\bibinfo {year} {1970})}\BibitemShut {NoStop}%
\bibitem [{\citenamefont {Black}\ \emph {et~al.}(2003)\citenamefont {Black},
  \citenamefont {Chan},\ and\ \citenamefont {Vuleti\ifmmode~\acute{c}\else
  \'{c}\fi{}}}]{PhysRevLett.91.203001}%
  \BibitemOpen
  \bibfield  {author} {\bibinfo {author} {\bibfnamefont {A.~T.}\ \bibnamefont
  {Black}}, \bibinfo {author} {\bibfnamefont {H.~W.}\ \bibnamefont {Chan}}, \
  and\ \bibinfo {author} {\bibfnamefont {V.}~\bibnamefont
  {Vuleti\ifmmode~\acute{c}\else \'{c}\fi{}}},\ }\href {\doibase
  10.1103/PhysRevLett.91.203001} {\bibfield  {journal} {\bibinfo  {journal}
  {Phys. Rev. Lett.}\ }\textbf {\bibinfo {volume} {91}},\ \bibinfo {pages}
  {203001} (\bibinfo {year} {2003})}\BibitemShut {NoStop}%
\bibitem [{\citenamefont {Bajcsy}\ \emph {et~al.}(2009)\citenamefont {Bajcsy},
  \citenamefont {Hofferberth}, \citenamefont {Balic}, \citenamefont {Peyronel},
  \citenamefont {Hafezi}, \citenamefont {Zibrov}, \citenamefont
  {Vuleti\ifmmode~\acute{c}\else \'{c}\fi{}},\ and\ \citenamefont
  {Lukin}}]{PhysRevLett.102.203902}%
  \BibitemOpen
  \bibfield  {author} {\bibinfo {author} {\bibfnamefont {M.}~\bibnamefont
  {Bajcsy}}, \bibinfo {author} {\bibfnamefont {S.}~\bibnamefont {Hofferberth}},
  \bibinfo {author} {\bibfnamefont {V.}~\bibnamefont {Balic}}, \bibinfo
  {author} {\bibfnamefont {T.}~\bibnamefont {Peyronel}}, \bibinfo {author}
  {\bibfnamefont {M.}~\bibnamefont {Hafezi}}, \bibinfo {author} {\bibfnamefont
  {A.~S.}\ \bibnamefont {Zibrov}}, \bibinfo {author} {\bibfnamefont
  {V.}~\bibnamefont {Vuleti\ifmmode~\acute{c}\else \'{c}\fi{}}}, \ and\
  \bibinfo {author} {\bibfnamefont {M.~D.}\ \bibnamefont {Lukin}},\ }\href
  {\doibase 10.1103/PhysRevLett.102.203902} {\bibfield  {journal} {\bibinfo
  {journal} {Phys. Rev. Lett.}\ }\textbf {\bibinfo {volume} {102}},\ \bibinfo
  {pages} {203902} (\bibinfo {year} {2009})}\BibitemShut {NoStop}%
\bibitem [{\citenamefont {Eisaman}\ \emph {et~al.}(2005)\citenamefont
  {Eisaman}, \citenamefont {Andre}, \citenamefont {Massou}, \citenamefont
  {Fleischhauer}, \citenamefont {Zibrov},\ and\ \citenamefont
  {Lukin}}]{Nature438}%
  \BibitemOpen
  \bibfield  {author} {\bibinfo {author} {\bibfnamefont {M.~D.}\ \bibnamefont
  {Eisaman}}, \bibinfo {author} {\bibfnamefont {A.}~\bibnamefont {Andre}},
  \bibinfo {author} {\bibfnamefont {F.}~\bibnamefont {Massou}}, \bibinfo
  {author} {\bibfnamefont {M.}~\bibnamefont {Fleischhauer}}, \bibinfo {author}
  {\bibfnamefont {A.~S.}\ \bibnamefont {Zibrov}}, \ and\ \bibinfo {author}
  {\bibfnamefont {M.~D.}\ \bibnamefont {Lukin}},\ }\href {\doibase
  10.1038/nature04327} {\bibfield  {journal} {\bibinfo  {journal} {Nature}\
  }\textbf {\bibinfo {volume} {438}},\ \bibinfo {pages} {837} (\bibinfo {year}
  {2005})}\BibitemShut {NoStop}%
\bibitem [{\citenamefont {Dudin}\ and\ \citenamefont
  {Kuzmich}(2012)}]{Dudin18052012}%
  \BibitemOpen
  \bibfield  {author} {\bibinfo {author} {\bibfnamefont {Y.~O.}\ \bibnamefont
  {Dudin}}\ and\ \bibinfo {author} {\bibfnamefont {A.}~\bibnamefont
  {Kuzmich}},\ }\href {\doibase 10.1126/science.1217901} {\bibfield  {journal}
  {\bibinfo  {journal} {Science}\ }\textbf {\bibinfo {volume} {336}},\ \bibinfo
  {pages} {887} (\bibinfo {year} {2012})}\BibitemShut {NoStop}%
\bibitem [{\citenamefont {Parigi}\ \emph {et~al.}(2012)\citenamefont {Parigi},
  \citenamefont {Bimbard}, \citenamefont {Stanojevic}, \citenamefont
  {Hilliard}, \citenamefont {Nogrette}, \citenamefont {Tualle-Brouri},
  \citenamefont {Ourjoumtsev},\ and\ \citenamefont
  {Grangier}}]{PhysRevLett.109.233602}%
  \BibitemOpen
  \bibfield  {author} {\bibinfo {author} {\bibfnamefont {V.}~\bibnamefont
  {Parigi}}, \bibinfo {author} {\bibfnamefont {E.}~\bibnamefont {Bimbard}},
  \bibinfo {author} {\bibfnamefont {J.}~\bibnamefont {Stanojevic}}, \bibinfo
  {author} {\bibfnamefont {A.~J.}\ \bibnamefont {Hilliard}}, \bibinfo {author}
  {\bibfnamefont {F.}~\bibnamefont {Nogrette}}, \bibinfo {author}
  {\bibfnamefont {R.}~\bibnamefont {Tualle-Brouri}}, \bibinfo {author}
  {\bibfnamefont {A.}~\bibnamefont {Ourjoumtsev}}, \ and\ \bibinfo {author}
  {\bibfnamefont {P.}~\bibnamefont {Grangier}},\ }\href {\doibase
  10.1103/PhysRevLett.109.233602} {\bibfield  {journal} {\bibinfo  {journal}
  {Phys. Rev. Lett.}\ }\textbf {\bibinfo {volume} {109}},\ \bibinfo {pages}
  {233602} (\bibinfo {year} {2012})}\BibitemShut {NoStop}%
\bibitem [{\citenamefont {Birnbaum}\ \emph {et~al.}(2005)\citenamefont
  {Birnbaum}, \citenamefont {Boca}, \citenamefont {Miller}, \citenamefont
  {Boozer}, \citenamefont {Northup},\ and\ \citenamefont {Kimble}}]{Nature436}%
  \BibitemOpen
  \bibfield  {author} {\bibinfo {author} {\bibfnamefont {K.~M.}\ \bibnamefont
  {Birnbaum}}, \bibinfo {author} {\bibfnamefont {A.}~\bibnamefont {Boca}},
  \bibinfo {author} {\bibfnamefont {R.}~\bibnamefont {Miller}}, \bibinfo
  {author} {\bibfnamefont {A.~D.}\ \bibnamefont {Boozer}}, \bibinfo {author}
  {\bibfnamefont {T.~E.}\ \bibnamefont {Northup}}, \ and\ \bibinfo {author}
  {\bibfnamefont {H.~J.}\ \bibnamefont {Kimble}},\ }\href {\doibase
  10.1038/nature03804} {\bibfield  {journal} {\bibinfo  {journal} {Nature}\
  }\textbf {\bibinfo {volume} {436}},\ \bibinfo {pages} {87} (\bibinfo {year}
  {2005})}\BibitemShut {NoStop}%
\bibitem [{\citenamefont {Tanji-Suzuki}\ \emph {et~al.}(2011)\citenamefont
  {Tanji-Suzuki}, \citenamefont {Chen}, \citenamefont {Landig}, \citenamefont
  {Simon},\ and\ \citenamefont {Vuleti\ifmmode~\acute{c}\else
  \'{c}\fi{}}}]{Tanji-Suzuki02092011}%
  \BibitemOpen
  \bibfield  {author} {\bibinfo {author} {\bibfnamefont {H.}~\bibnamefont
  {Tanji-Suzuki}}, \bibinfo {author} {\bibfnamefont {W.}~\bibnamefont {Chen}},
  \bibinfo {author} {\bibfnamefont {R.}~\bibnamefont {Landig}}, \bibinfo
  {author} {\bibfnamefont {J.}~\bibnamefont {Simon}}, \ and\ \bibinfo {author}
  {\bibfnamefont {V.}~\bibnamefont {Vuleti\ifmmode~\acute{c}\else
  \'{c}\fi{}}},\ }\href {\doibase 10.1126/science.1208066} {\bibfield
  {journal} {\bibinfo  {journal} {Science}\ }\textbf {\bibinfo {volume}
  {333}},\ \bibinfo {pages} {1266} (\bibinfo {year} {2011})}\BibitemShut
  {NoStop}%
\bibitem [{\citenamefont {Peyronel}\ \emph {et~al.}(2012)\citenamefont
  {Peyronel}, \citenamefont {Firstenberg}, \citenamefont {Liang}, \citenamefont
  {Hofferberth}, \citenamefont {Gorshkov}, \citenamefont {Pohl}, \citenamefont
  {Lukin},\ and\ \citenamefont {Vuleti\ifmmode~\acute{c}\else
  \'{c}\fi{}}}]{Peyronel}%
  \BibitemOpen
  \bibfield  {author} {\bibinfo {author} {\bibfnamefont {T.}~\bibnamefont
  {Peyronel}}, \bibinfo {author} {\bibfnamefont {O.}~\bibnamefont
  {Firstenberg}}, \bibinfo {author} {\bibfnamefont {Q.-Y.}\ \bibnamefont
  {Liang}}, \bibinfo {author} {\bibfnamefont {S.}~\bibnamefont {Hofferberth}},
  \bibinfo {author} {\bibfnamefont {A.~V.}\ \bibnamefont {Gorshkov}}, \bibinfo
  {author} {\bibfnamefont {T.}~\bibnamefont {Pohl}}, \bibinfo {author}
  {\bibfnamefont {M.~D.}\ \bibnamefont {Lukin}}, \ and\ \bibinfo {author}
  {\bibfnamefont {V.}~\bibnamefont {Vuleti\ifmmode~\acute{c}\else
  \'{c}\fi{}}},\ }\href@noop {} {\bibfield  {journal} {\bibinfo  {journal}
  {Nature}\ }\textbf {\bibinfo {volume} {488}},\ \bibinfo {pages} {57}
  (\bibinfo {year} {2012})}\BibitemShut {NoStop}%
\bibitem [{\citenamefont {Baur}\ \emph {et~al.}(2014)\citenamefont {Baur},
  \citenamefont {Tiarks}, \citenamefont {Rempe},\ and\ \citenamefont
  {D\"urr}}]{PhysRevLett.112.073901}%
  \BibitemOpen
  \bibfield  {author} {\bibinfo {author} {\bibfnamefont {S.}~\bibnamefont
  {Baur}}, \bibinfo {author} {\bibfnamefont {D.}~\bibnamefont {Tiarks}},
  \bibinfo {author} {\bibfnamefont {G.}~\bibnamefont {Rempe}}, \ and\ \bibinfo
  {author} {\bibfnamefont {S.}~\bibnamefont {D\"urr}},\ }\href {\doibase
  10.1103/PhysRevLett.112.073901} {\bibfield  {journal} {\bibinfo  {journal}
  {Phys. Rev. Lett.}\ }\textbf {\bibinfo {volume} {112}},\ \bibinfo {pages}
  {073901} (\bibinfo {year} {2014})}\BibitemShut {NoStop}%
\bibitem [{\citenamefont {Harris}\ and\ \citenamefont {Hau}(1999)}]{Harris}%
  \BibitemOpen
  \bibfield  {author} {\bibinfo {author} {\bibfnamefont {S.~E.}\ \bibnamefont
  {Harris}}\ and\ \bibinfo {author} {\bibfnamefont {L.~V.}\ \bibnamefont
  {Hau}},\ }\href@noop {} {\bibfield  {journal} {\bibinfo  {journal} {Phys.
  Rev. Lett.}\ }\textbf {\bibinfo {volume} {82}},\ \bibinfo {pages} {23}
  (\bibinfo {year} {1999})}\BibitemShut {NoStop}%
\bibitem [{\citenamefont {Dawes}\ \emph {et~al.}(2010)\citenamefont {Dawes},
  \citenamefont {Gauthier}, \citenamefont {Schumacher}, \citenamefont {Kwong},
  \citenamefont {Binder},\ and\ \citenamefont {Smirl}}]{Dawes}%
  \BibitemOpen
  \bibfield  {author} {\bibinfo {author} {\bibfnamefont {A.~M.~C.}\
  \bibnamefont {Dawes}}, \bibinfo {author} {\bibfnamefont {D.~J.}\ \bibnamefont
  {Gauthier}}, \bibinfo {author} {\bibfnamefont {S.}~\bibnamefont
  {Schumacher}}, \bibinfo {author} {\bibfnamefont {N.~H.}\ \bibnamefont
  {Kwong}}, \bibinfo {author} {\bibfnamefont {R.}~\bibnamefont {Binder}}, \
  and\ \bibinfo {author} {\bibfnamefont {A.~L.}\ \bibnamefont {Smirl}},\ }\href
  {\doibase 10.1002/lpor.200810067} {\bibfield  {journal} {\bibinfo  {journal}
  {Laser \& Photonics Reviews}\ }\textbf {\bibinfo {volume} {4}},\ \bibinfo
  {pages} {221} (\bibinfo {year} {2010})}\BibitemShut {NoStop}%
\bibitem [{\citenamefont {Ritsch}\ \emph {et~al.}(2013)\citenamefont {Ritsch},
  \citenamefont {Domokos}, \citenamefont {Brennecke},\ and\ \citenamefont
  {Esslinger}}]{Ritsch}%
  \BibitemOpen
  \bibfield  {author} {\bibinfo {author} {\bibfnamefont {H.}~\bibnamefont
  {Ritsch}}, \bibinfo {author} {\bibfnamefont {P.}~\bibnamefont {Domokos}},
  \bibinfo {author} {\bibfnamefont {F.}~\bibnamefont {Brennecke}}, \ and\
  \bibinfo {author} {\bibfnamefont {T.}~\bibnamefont {Esslinger}},\ }\href@noop
  {} {\bibfield  {journal} {\bibinfo  {journal} {Rev. Mod. Phys.}\ }\textbf
  {\bibinfo {volume} {85}},\ \bibinfo {pages} {2} (\bibinfo {year}
  {2013})}\BibitemShut {NoStop}%
\bibitem [{\citenamefont {Deutsch}\ \emph {et~al.}(1995)\citenamefont
  {Deutsch}, \citenamefont {Spreeuw}, \citenamefont {Rolston},\ and\
  \citenamefont {Phillips}}]{Deutsch}%
  \BibitemOpen
  \bibfield  {author} {\bibinfo {author} {\bibfnamefont {I.~H.}\ \bibnamefont
  {Deutsch}}, \bibinfo {author} {\bibfnamefont {R.~J.~C.}\ \bibnamefont
  {Spreeuw}}, \bibinfo {author} {\bibfnamefont {S.~L.}\ \bibnamefont
  {Rolston}}, \ and\ \bibinfo {author} {\bibfnamefont {W.~D.}\ \bibnamefont
  {Phillips}},\ }\href@noop {} {\bibfield  {journal} {\bibinfo  {journal}
  {Phys. Rev. A}\ }\textbf {\bibinfo {volume} {52}},\ \bibinfo {pages} {2}
  (\bibinfo {year} {1995})}\BibitemShut {NoStop}%
\bibitem [{\citenamefont {Asboth}\ \emph {et~al.}(2008)\citenamefont {Asboth},
  \citenamefont {Ritsch},\ and\ \citenamefont {Domokos}}]{asboth2008}%
  \BibitemOpen
  \bibfield  {author} {\bibinfo {author} {\bibfnamefont {J.~K.}\ \bibnamefont
  {Asboth}}, \bibinfo {author} {\bibfnamefont {H.}~\bibnamefont {Ritsch}}, \
  and\ \bibinfo {author} {\bibfnamefont {P.}~\bibnamefont {Domokos}},\
  }\href@noop {} {\bibfield  {journal} {\bibinfo  {journal} {Phys. Rev. A}\
  }\textbf {\bibinfo {volume} {77}},\ \bibinfo {pages} {063424} (\bibinfo
  {year} {2008})}\BibitemShut {NoStop}%
\bibitem [{\citenamefont {Asboth}\ \emph {et~al.}(2007)\citenamefont {Asboth},
  \citenamefont {Ritsch},\ and\ \citenamefont {Domokos}}]{asboth2007}%
  \BibitemOpen
  \bibfield  {author} {\bibinfo {author} {\bibfnamefont {J.~K.}\ \bibnamefont
  {Asboth}}, \bibinfo {author} {\bibfnamefont {H.}~\bibnamefont {Ritsch}}, \
  and\ \bibinfo {author} {\bibfnamefont {P.}~\bibnamefont {Domokos}},\
  }\href@noop {} {\bibfield  {journal} {\bibinfo  {journal} {Phys. Rev. Lett.}\
  }\textbf {\bibinfo {volume} {98}},\ \bibinfo {pages} {203008} (\bibinfo
  {year} {2007})}\BibitemShut {NoStop}%
\bibitem [{\citenamefont {Els\"{a}sser}\ \emph {et~al.}(2004)\citenamefont
  {Els\"{a}sser}, \citenamefont {Nagorny},\ and\ \citenamefont
  {Hemmerich}}]{Elsasser}%
  \BibitemOpen
  \bibfield  {author} {\bibinfo {author} {\bibfnamefont {T.}~\bibnamefont
  {Els\"{a}sser}}, \bibinfo {author} {\bibfnamefont {B.}~\bibnamefont
  {Nagorny}}, \ and\ \bibinfo {author} {\bibfnamefont {A.}~\bibnamefont
  {Hemmerich}},\ }\href@noop {} {\bibfield  {journal} {\bibinfo  {journal}
  {Phys. Rev. A}\ }\textbf {\bibinfo {volume} {69}},\ \bibinfo {pages} {033403}
  (\bibinfo {year} {2004})}\BibitemShut {NoStop}%
\bibitem [{\citenamefont {Asboth}\ \emph {et~al.}(2005)\citenamefont {Asboth},
  \citenamefont {Domokos}, \citenamefont {Ritsch},\ and\ \citenamefont
  {Vukics}}]{asboth2005}%
  \BibitemOpen
  \bibfield  {author} {\bibinfo {author} {\bibfnamefont {J.~K.}\ \bibnamefont
  {Asboth}}, \bibinfo {author} {\bibfnamefont {P.}~\bibnamefont {Domokos}},
  \bibinfo {author} {\bibfnamefont {H.}~\bibnamefont {Ritsch}}, \ and\ \bibinfo
  {author} {\bibfnamefont {A.}~\bibnamefont {Vukics}},\ }\href@noop {}
  {\bibfield  {journal} {\bibinfo  {journal} {Phys. Rev. A}\ }\textbf {\bibinfo
  {volume} {72}},\ \bibinfo {pages} {053417} (\bibinfo {year}
  {2005})}\BibitemShut {NoStop}%
\bibitem [{\citenamefont {Petrosyan}(2007)}]{Petrosyan}%
  \BibitemOpen
  \bibfield  {author} {\bibinfo {author} {\bibfnamefont {D.}~\bibnamefont
  {Petrosyan}},\ }\href {\doibase 10.1103/PhysRevA.76.053823} {\bibfield
  {journal} {\bibinfo  {journal} {Phys. Rev. A}\ }\textbf {\bibinfo {volume}
  {76}},\ \bibinfo {pages} {053823} (\bibinfo {year} {2007})}\BibitemShut
  {NoStop}%
\bibitem [{\citenamefont {Horsely}\ \emph {et~al.}(2008)\citenamefont
  {Horsely}, \citenamefont {Wu}, \citenamefont {Artoni},\ and\ \citenamefont
  {La~Rocca}}]{Wu}%
  \BibitemOpen
  \bibfield  {author} {\bibinfo {author} {\bibfnamefont {S.~A.~R.}\
  \bibnamefont {Horsely}}, \bibinfo {author} {\bibfnamefont {J.-H.}\
  \bibnamefont {Wu}}, \bibinfo {author} {\bibfnamefont {M.}~\bibnamefont
  {Artoni}}, \ and\ \bibinfo {author} {\bibfnamefont {G.~C.}\ \bibnamefont
  {La~Rocca}},\ }\href@noop {} {\bibfield  {journal} {\bibinfo  {journal} {J.
  Opt. Soc. Am. B}\ }\textbf {\bibinfo {volume} {25}},\ \bibinfo {pages} {11}
  (\bibinfo {year} {2008})}\BibitemShut {NoStop}%
\bibitem [{\citenamefont {Nunn}\ \emph {et~al.}(2010)\citenamefont {Nunn},
  \citenamefont {Dorner}, \citenamefont {Michelberger}, \citenamefont {Reim},
  \citenamefont {Lee}, \citenamefont {Langford}, \citenamefont {Walmsley},\
  and\ \citenamefont {Jaksch}}]{Nunn}%
  \BibitemOpen
  \bibfield  {author} {\bibinfo {author} {\bibfnamefont {J.}~\bibnamefont
  {Nunn}}, \bibinfo {author} {\bibfnamefont {U.}~\bibnamefont {Dorner}},
  \bibinfo {author} {\bibfnamefont {P.}~\bibnamefont {Michelberger}}, \bibinfo
  {author} {\bibfnamefont {K.~F.}\ \bibnamefont {Reim}}, \bibinfo {author}
  {\bibfnamefont {K.~C.}\ \bibnamefont {Lee}}, \bibinfo {author} {\bibfnamefont
  {N.~K.}\ \bibnamefont {Langford}}, \bibinfo {author} {\bibfnamefont {I.~A.}\
  \bibnamefont {Walmsley}}, \ and\ \bibinfo {author} {\bibfnamefont
  {D.}~\bibnamefont {Jaksch}},\ }\href@noop {} {\bibfield  {journal} {\bibinfo
  {journal} {Phys. Rev. A}\ }\textbf {\bibinfo {volume} {82}},\ \bibinfo
  {pages} {022327} (\bibinfo {year} {2010})}\BibitemShut {NoStop}%
\bibitem [{\citenamefont {Schilke}\ \emph
  {et~al.}(2012{\natexlab{a}})\citenamefont {Schilke}, \citenamefont
  {Zimmermann},\ and\ \citenamefont {Guerin}}]{Schilke}%
  \BibitemOpen
  \bibfield  {author} {\bibinfo {author} {\bibfnamefont {A.}~\bibnamefont
  {Schilke}}, \bibinfo {author} {\bibfnamefont {C.}~\bibnamefont {Zimmermann}},
  \ and\ \bibinfo {author} {\bibfnamefont {W.}~\bibnamefont {Guerin}},\
  }\href@noop {} {\bibfield  {journal} {\bibinfo  {journal} {Phys. Rev. A}\
  }\textbf {\bibinfo {volume} {86}},\ \bibinfo {pages} {023809} (\bibinfo
  {year} {2012}{\natexlab{a}})}\BibitemShut {NoStop}%
\bibitem [{\citenamefont {Schilke}\ \emph
  {et~al.}(2012{\natexlab{b}})\citenamefont {Schilke}, \citenamefont
  {Zimmerman}, \citenamefont {Courteille},\ and\ \citenamefont
  {Guerin}}]{SchilkeExpTransverse}%
  \BibitemOpen
  \bibfield  {author} {\bibinfo {author} {\bibfnamefont {A.}~\bibnamefont
  {Schilke}}, \bibinfo {author} {\bibfnamefont {C.}~\bibnamefont {Zimmerman}},
  \bibinfo {author} {\bibfnamefont {P.~W.}\ \bibnamefont {Courteille}}, \ and\
  \bibinfo {author} {\bibfnamefont {W.}~\bibnamefont {Guerin}},\ }\href
  {\doibase 10.1038/nphoton.2011.320} {\bibfield  {journal} {\bibinfo
  {journal} {Nat. Photon.}\ }\textbf {\bibinfo {volume} {6}},\ \bibinfo {pages}
  {101} (\bibinfo {year} {2012}{\natexlab{b}})}\BibitemShut {NoStop}%
\bibitem [{\citenamefont {Greenberg}\ \emph {et~al.}(2011)\citenamefont
  {Greenberg}, \citenamefont {Schmittberger},\ and\ \citenamefont
  {Gauthier}}]{GreenbergOptExp}%
  \BibitemOpen
  \bibfield  {author} {\bibinfo {author} {\bibfnamefont {J.~A.}\ \bibnamefont
  {Greenberg}}, \bibinfo {author} {\bibfnamefont {B.~L.}\ \bibnamefont
  {Schmittberger}}, \ and\ \bibinfo {author} {\bibfnamefont {D.~J.}\
  \bibnamefont {Gauthier}},\ }\href@noop {} {\bibfield  {journal} {\bibinfo
  {journal} {Opt. Express}\ }\textbf {\bibinfo {volume} {19}},\ \bibinfo
  {pages} {22535} (\bibinfo {year} {2011})}\BibitemShut {NoStop}%
\bibitem [{\citenamefont {Labeyrie}\ \emph {et~al.}(2014)\citenamefont
  {Labeyrie}, \citenamefont {Tesio}, \citenamefont {Gomes}, \citenamefont
  {Oppo}, \citenamefont {Firth}, \citenamefont {Robb}, \citenamefont {Arnold},
  \citenamefont {Kaiser},\ and\ \citenamefont {Ackemann}}]{Labeyrie}%
  \BibitemOpen
  \bibfield  {author} {\bibinfo {author} {\bibfnamefont {G.}~\bibnamefont
  {Labeyrie}}, \bibinfo {author} {\bibfnamefont {E.}~\bibnamefont {Tesio}},
  \bibinfo {author} {\bibfnamefont {P.~M.}\ \bibnamefont {Gomes}}, \bibinfo
  {author} {\bibfnamefont {G.-L.}\ \bibnamefont {Oppo}}, \bibinfo {author}
  {\bibfnamefont {W.~J.}\ \bibnamefont {Firth}}, \bibinfo {author}
  {\bibfnamefont {G.~R.~M.}\ \bibnamefont {Robb}}, \bibinfo {author}
  {\bibfnamefont {A.~S.}\ \bibnamefont {Arnold}}, \bibinfo {author}
  {\bibfnamefont {R.}~\bibnamefont {Kaiser}}, \ and\ \bibinfo {author}
  {\bibfnamefont {T.}~\bibnamefont {Ackemann}},\ }\href {\doibase
  10.1038/nphoton.2014.52} {\bibfield  {journal} {\bibinfo  {journal} {Nat.
  Photon.}\ }\textbf {\bibinfo {volume} {8}},\ \bibinfo {pages} {321} (\bibinfo
  {year} {2014})}\BibitemShut {NoStop}%
\bibitem [{\citenamefont {Birkl}\ \emph {et~al.}(1995)\citenamefont {Birkl},
  \citenamefont {Gatzke}, \citenamefont {Deutsch}, \citenamefont {Rolston},\
  and\ \citenamefont {Phillips}}]{PhysRevLett.75.2823}%
  \BibitemOpen
  \bibfield  {author} {\bibinfo {author} {\bibfnamefont {G.}~\bibnamefont
  {Birkl}}, \bibinfo {author} {\bibfnamefont {M.}~\bibnamefont {Gatzke}},
  \bibinfo {author} {\bibfnamefont {I.~H.}\ \bibnamefont {Deutsch}}, \bibinfo
  {author} {\bibfnamefont {S.~L.}\ \bibnamefont {Rolston}}, \ and\ \bibinfo
  {author} {\bibfnamefont {W.~D.}\ \bibnamefont {Phillips}},\ }\href {\doibase
  10.1103/PhysRevLett.75.2823} {\bibfield  {journal} {\bibinfo  {journal}
  {Phys. Rev. Lett.}\ }\textbf {\bibinfo {volume} {75}},\ \bibinfo {pages}
  {2823} (\bibinfo {year} {1995})}\BibitemShut {NoStop}%
\bibitem [{\citenamefont {Weidem\"uller}\ \emph {et~al.}(1995)\citenamefont
  {Weidem\"uller}, \citenamefont {Hemmerich}, \citenamefont {G\"orlitz},
  \citenamefont {Esslinger},\ and\ \citenamefont
  {H\"ansch}}]{PhysRevLett.75.4583}%
  \BibitemOpen
  \bibfield  {author} {\bibinfo {author} {\bibfnamefont {M.}~\bibnamefont
  {Weidem\"uller}}, \bibinfo {author} {\bibfnamefont {A.}~\bibnamefont
  {Hemmerich}}, \bibinfo {author} {\bibfnamefont {A.}~\bibnamefont
  {G\"orlitz}}, \bibinfo {author} {\bibfnamefont {T.}~\bibnamefont
  {Esslinger}}, \ and\ \bibinfo {author} {\bibfnamefont {T.~W.}\ \bibnamefont
  {H\"ansch}},\ }\href {\doibase 10.1103/PhysRevLett.75.4583} {\bibfield
  {journal} {\bibinfo  {journal} {Phys. Rev. Lett.}\ }\textbf {\bibinfo
  {volume} {75}},\ \bibinfo {pages} {4583} (\bibinfo {year}
  {1995})}\BibitemShut {NoStop}%
\bibitem [{\citenamefont {Muradyan}\ \emph {et~al.}(2005)\citenamefont
  {Muradyan}, \citenamefont {Wang}, \citenamefont {Williams},\ and\
  \citenamefont {Saffman}}]{Muradyan}%
  \BibitemOpen
  \bibfield  {author} {\bibinfo {author} {\bibfnamefont {G.}~\bibnamefont
  {Muradyan}}, \bibinfo {author} {\bibfnamefont {Y.}~\bibnamefont {Wang}},
  \bibinfo {author} {\bibfnamefont {W.}~\bibnamefont {Williams}}, \ and\
  \bibinfo {author} {\bibfnamefont {M.}~\bibnamefont {Saffman}},\ }\href@noop
  {} {\bibfield  {journal} {\bibinfo  {journal} {Nonlinear Guided Waves and
  Their Applications, Technical Digest (CD), Optical Society of America, paper
  ThB29}\ } (\bibinfo {year} {2005})}\BibitemShut {NoStop}%
\bibitem [{\citenamefont {Saffman}\ and\ \citenamefont {Wang}(2008)}]{Wang}%
  \BibitemOpen
  \bibfield  {author} {\bibinfo {author} {\bibfnamefont {M.}~\bibnamefont
  {Saffman}}\ and\ \bibinfo {author} {\bibfnamefont {Y.}~\bibnamefont {Wang}},\
  }\href@noop {} {\bibfield  {journal} {\bibinfo  {journal} {Lect. Notes
  Phys.}\ }\textbf {\bibinfo {volume} {751}},\ \bibinfo {pages} {361} (\bibinfo
  {year} {2008})}\BibitemShut {NoStop}%
\bibitem [{\citenamefont {Greenberg}\ and\ \citenamefont
  {Gauthier}(2012{\natexlab{a}})}]{GreenbergPRA}%
  \BibitemOpen
  \bibfield  {author} {\bibinfo {author} {\bibfnamefont {J.~A.}\ \bibnamefont
  {Greenberg}}\ and\ \bibinfo {author} {\bibfnamefont {D.~J.}\ \bibnamefont
  {Gauthier}},\ }\href@noop {} {\bibfield  {journal} {\bibinfo  {journal}
  {Phys. Rev. A}\ }\textbf {\bibinfo {volume} {86}},\ \bibinfo {pages} {013823}
  (\bibinfo {year} {2012}{\natexlab{a}})}\BibitemShut {NoStop}%
\bibitem [{\citenamefont {Gattobigio}\ \emph {et~al.}(2006)\citenamefont
  {Gattobigio}, \citenamefont {Michaud}, \citenamefont {Javaloyes},
  \citenamefont {Tabosa},\ and\ \citenamefont {Kaiser}}]{Gattobigio}%
  \BibitemOpen
  \bibfield  {author} {\bibinfo {author} {\bibfnamefont {G.~L.}\ \bibnamefont
  {Gattobigio}}, \bibinfo {author} {\bibfnamefont {F.}~\bibnamefont {Michaud}},
  \bibinfo {author} {\bibfnamefont {J.}~\bibnamefont {Javaloyes}}, \bibinfo
  {author} {\bibfnamefont {J.~W.~R.}\ \bibnamefont {Tabosa}}, \ and\ \bibinfo
  {author} {\bibfnamefont {R.}~\bibnamefont {Kaiser}},\ }\href@noop {}
  {\bibfield  {journal} {\bibinfo  {journal} {Phys. Rev. A}\ }\textbf {\bibinfo
  {volume} {74}},\ \bibinfo {pages} {043407} (\bibinfo {year}
  {2006})}\BibitemShut {NoStop}%
\bibitem [{\citenamefont {Arnold}\ \emph {et~al.}(2012)\citenamefont {Arnold},
  \citenamefont {Baden},\ and\ \citenamefont {Barrett}}]{Arnold}%
  \BibitemOpen
  \bibfield  {author} {\bibinfo {author} {\bibfnamefont {K.~J.}\ \bibnamefont
  {Arnold}}, \bibinfo {author} {\bibfnamefont {M.~P.}\ \bibnamefont {Baden}}, \
  and\ \bibinfo {author} {\bibfnamefont {M.~D.}\ \bibnamefont {Barrett}},\
  }\href@noop {} {\bibfield  {journal} {\bibinfo  {journal} {Phys. Rev. Lett.}\
  }\textbf {\bibinfo {volume} {109}},\ \bibinfo {pages} {153002} (\bibinfo
  {year} {2012})}\BibitemShut {NoStop}%
\bibitem [{\citenamefont {Deng}\ \emph {et~al.}(2010)\citenamefont {Deng},
  \citenamefont {Hagley}, \citenamefont {Cao}, \citenamefont {Wang},
  \citenamefont {Luo}, \citenamefont {Wang}, \citenamefont {Payne},
  \citenamefont {Yang}, \citenamefont {Zhou}, \citenamefont {Chen},\ and\
  \citenamefont {Zhan}}]{Deng}%
  \BibitemOpen
  \bibfield  {author} {\bibinfo {author} {\bibfnamefont {L.}~\bibnamefont
  {Deng}}, \bibinfo {author} {\bibfnamefont {E.~W.}\ \bibnamefont {Hagley}},
  \bibinfo {author} {\bibfnamefont {Q.}~\bibnamefont {Cao}}, \bibinfo {author}
  {\bibfnamefont {X.}~\bibnamefont {Wang}}, \bibinfo {author} {\bibfnamefont
  {X.}~\bibnamefont {Luo}}, \bibinfo {author} {\bibfnamefont {R.}~\bibnamefont
  {Wang}}, \bibinfo {author} {\bibfnamefont {M.~G.}\ \bibnamefont {Payne}},
  \bibinfo {author} {\bibfnamefont {F.}~\bibnamefont {Yang}}, \bibinfo {author}
  {\bibfnamefont {X.}~\bibnamefont {Zhou}}, \bibinfo {author} {\bibfnamefont
  {X.}~\bibnamefont {Chen}}, \ and\ \bibinfo {author} {\bibfnamefont
  {M.}~\bibnamefont {Zhan}},\ }\href@noop {} {\bibfield  {journal} {\bibinfo
  {journal} {Phys. Rev. Lett.}\ }\textbf {\bibinfo {volume} {105}},\ \bibinfo
  {pages} {220404} (\bibinfo {year} {2010})}\BibitemShut {NoStop}%
\bibitem [{\citenamefont {Boyd}(2008)}]{Boyd}%
  \BibitemOpen
  \bibfield  {author} {\bibinfo {author} {\bibfnamefont {R.~W.}\ \bibnamefont
  {Boyd}},\ }\href@noop {} {\emph {\bibinfo {title} {Nonlinear Optics, 3rd
  Ed.}}}\ (\bibinfo  {publisher} {Academic Press},\ \bibinfo {year}
  {2008})\BibitemShut {NoStop}%
\bibitem [{\citenamefont {Tesio}\ \emph {et~al.}(2012)\citenamefont {Tesio},
  \citenamefont {Robb}, \citenamefont {Ackemann}, \citenamefont {Firth},\ and\
  \citenamefont {Oppo}}]{Tesio}%
  \BibitemOpen
  \bibfield  {author} {\bibinfo {author} {\bibfnamefont {E.}~\bibnamefont
  {Tesio}}, \bibinfo {author} {\bibfnamefont {G.~R.~M.}\ \bibnamefont {Robb}},
  \bibinfo {author} {\bibfnamefont {T.}~\bibnamefont {Ackemann}}, \bibinfo
  {author} {\bibfnamefont {W.~J.}\ \bibnamefont {Firth}}, \ and\ \bibinfo
  {author} {\bibfnamefont {G.-L.}\ \bibnamefont {Oppo}},\ }\href@noop {}
  {\bibfield  {journal} {\bibinfo  {journal} {Phys. Rev. A}\ }\textbf {\bibinfo
  {volume} {86}},\ \bibinfo {pages} {031801(R)} (\bibinfo {year}
  {2012})}\BibitemShut {NoStop}%
\bibitem [{\citenamefont {Bonifacio}\ \emph {et~al.}(1994)\citenamefont
  {Bonifacio}, \citenamefont {De~Salvo}, \citenamefont {Narducci},\ and\
  \citenamefont {D'Angelo}}]{PhysRevA.50.1716}%
  \BibitemOpen
  \bibfield  {author} {\bibinfo {author} {\bibfnamefont {R.}~\bibnamefont
  {Bonifacio}}, \bibinfo {author} {\bibfnamefont {L.}~\bibnamefont {De~Salvo}},
  \bibinfo {author} {\bibfnamefont {L.~M.}\ \bibnamefont {Narducci}}, \ and\
  \bibinfo {author} {\bibfnamefont {E.~J.}\ \bibnamefont {D'Angelo}},\ }\href
  {\doibase 10.1103/PhysRevA.50.1716} {\bibfield  {journal} {\bibinfo
  {journal} {Phys. Rev. A}\ }\textbf {\bibinfo {volume} {50}},\ \bibinfo
  {pages} {1716} (\bibinfo {year} {1994})}\BibitemShut {NoStop}%
\bibitem [{\citenamefont {Chiao}\ \emph {et~al.}(1966)\citenamefont {Chiao},
  \citenamefont {Kelley},\ and\ \citenamefont {Garmie}}]{Chiao}%
  \BibitemOpen
  \bibfield  {author} {\bibinfo {author} {\bibfnamefont {R.~Y.}\ \bibnamefont
  {Chiao}}, \bibinfo {author} {\bibfnamefont {P.~L.}\ \bibnamefont {Kelley}}, \
  and\ \bibinfo {author} {\bibfnamefont {E.}~\bibnamefont {Garmie}},\
  }\href@noop {} {\bibfield  {journal} {\bibinfo  {journal} {Phys. Rev. Lett.}\
  }\textbf {\bibinfo {volume} {17}},\ \bibinfo {pages} {1158} (\bibinfo {year}
  {1966})}\BibitemShut {NoStop}%
\bibitem [{\citenamefont {Schmidt}\ and\ \citenamefont
  {A.~Imamo\v{g}lu}(1996)}]{Schmidt1996}%
  \BibitemOpen
  \bibfield  {author} {\bibinfo {author} {\bibfnamefont {H.}~\bibnamefont
  {Schmidt}}\ and\ \bibinfo {author} {\bibfnamefont {A.}~\bibnamefont
  {A.~Imamo\v{g}lu}},\ }\href {\doibase 10.1364/OL.21.001936} {\bibfield
  {journal} {\bibinfo  {journal} {Opt. Lett.}\ }\textbf {\bibinfo {volume}
  {21}},\ \bibinfo {pages} {1936} (\bibinfo {year} {1996})}\BibitemShut
  {NoStop}%
\bibitem [{\citenamefont {Firth}\ \emph {et~al.}(1990)\citenamefont {Firth},
  \citenamefont {Fitzgerald},\ and\ \citenamefont {Par\'{e}}}]{Firth90}%
  \BibitemOpen
  \bibfield  {author} {\bibinfo {author} {\bibfnamefont {W.~J.}\ \bibnamefont
  {Firth}}, \bibinfo {author} {\bibfnamefont {A.}~\bibnamefont {Fitzgerald}}, \
  and\ \bibinfo {author} {\bibfnamefont {C.}~\bibnamefont {Par\'{e}}},\ }\href
  {\doibase 10.1364/JOSAB.7.001087} {\bibfield  {journal} {\bibinfo  {journal}
  {J. Opt. Soc. Am. B}\ }\textbf {\bibinfo {volume} {7}},\ \bibinfo {pages}
  {1087} (\bibinfo {year} {1990})}\BibitemShut {NoStop}%
\bibitem [{\citenamefont {Greenberg}\ and\ \citenamefont
  {Gauthier}(2012{\natexlab{b}})}]{GreenbergEPL}%
  \BibitemOpen
  \bibfield  {author} {\bibinfo {author} {\bibfnamefont {J.~A.}\ \bibnamefont
  {Greenberg}}\ and\ \bibinfo {author} {\bibfnamefont {D.~J.}\ \bibnamefont
  {Gauthier}},\ }\href@noop {} {\bibfield  {journal} {\bibinfo  {journal} {Eur.
  Phys. Lett.}\ }\textbf {\bibinfo {volume} {98}},\ \bibinfo {pages} {24001}
  (\bibinfo {year} {2012}{\natexlab{b}})}\BibitemShut {NoStop}%
\end{thebibliography}%
\end{document}